\documentclass[aps,prb,twocolumn,nobalancelastpage,%
footinbib,%
noshowpacs,preprintnumbers,superscriptaddress%
]{revtex4-1}
\usepackage{amsmath,amssymb,amsfonts,fixmath,siunitx}
\usepackage{bm,graphicx,color}
\usepackage[pdftex,colorlinks=true,
					 bookmarksnumbered,bookmarksopen=true,bookmarksopenlevel=-1]{hyperref}

\def\dag{^{\dagger}}

\newcommand{\bra}[1]{\langle #1 \vert} 
\newcommand{\ket}[1]{\vert #1 \rangle} 
\newcommand{\ip}[2]{\langle #1 \vert #2 \rangle}  

\def\dna{\downarrow}
\def\upa{\uparrow}

\def\d{\delta}
\def\D{\Delta}
\def\e{\epsilon}
\def\ve{\varepsilon}

\def\s{\sigma}

\def\w{\omega}

\newcommand{\abs}[1]{\left\vert #1\right\vert}
\newcommand{\appref}[1]{Appendix \ref{#1}}

\def\const{\tx{const}}
\def\cp{\citep}

\def\ct{\citet}

\def\eg{e.g.\ }
\def\Eqref{Eq.~\eqref}

\newcommand{\expec}[1]{\langle #1 \rangle}

\newcommand{\figref}[1]{Fig.~\ref{#1}}
\newcommand{\figsref}[1]{Figs.~\ref{#1}}

\def\ie{i.e.\ }
\newcommand{\ig}[2]{\includegraphics[width=#1\tw,type=#2,ext=.#2,read=.#2]}

\def\lb{\label}

\def\non{\nonumber}

\newcommand{\secref}[1]{Sec.~\ref{#1}}

\renewcommand{\th}[1]{^\tx{#1}}

\def\tit{\textit}

\newcommand{\tl}[1]{_\tx{#1}}

\def\tw{\textwidth}
\def\tx{\text}

\def\wt{\widetilde}

\newcommand{\eq}[1]{\begin{align} #1 \end{align}}
\newcommand{\eqs}[1]{\begin{subequations}\begin{align} #1 \end{align}\end{subequations}}

\def\bar{\begin{array}}
\def\ear{\end{array}}

\def\bce{\begin{center}}
\def\ece{\end{center}}

\def\ben{\begin{enumerate}}
\def\een{\end{enumerate}}

\def\bfi{\begin{figure}[!tbh]\setcapindent{0em}\centering}
\newcommand{\bfiO}[1]{\begin{figure}[#1]\setcapindent{0em}\centering}
\def\efi{\end{figure}}

\def\bit{\begin{itemize}}
\def\eit{\end{itemize}}

\def\bqu{\begin{quote}}
\def\equ{\end{quote}}

\newcommand{\btblO}[2]{\begin{table}[#2]\setcapindent{0em}\centering\begin{minipage}{#1\tw}}
\newcommand{\btbl}[1]{\begin{table}[!tbh]\setcapindent{0em}\centering\begin{minipage}{#1\tw}}
\def\etbl{\end{minipage}\end{table}}

\def\btbr{\centering\vspace{.4em}\begin{tabular}}
\def\etbr{\end{tabular}}

\def\bseq{\begin{subequations}}
\def\eseq{\end{subequations}}

\def\bve{\begin{array}{l}}
\def\eve{\end{array}{l}}

\newcommand{\igS}[3]{%
\hspace{-.025\tw}\begin{minipage}[t]{#1\tw}\vspace{0em}\includegraphics[width=\linewidth,type=#2,ext=.#2,read=.#2]{#3}\end{minipage}%
\hspace{.03\tw}\begin{minipage}[t]{1\tw-#1\tw-.05\tw}}

\newcommand{\lbS}[1]{\lb{#1}\end{minipage}}

\makeatletter
  \newcommand{\myhref}[2]{\hyper@linkurl{#2}{#1}}
\makeatother

\makeatletter
\newcommand*\if@single[3]{%
  \setbox0\hbox{${\mathaccent"0362{#1}}^H$}%
  \setbox2\hbox{${\mathaccent"0362{\kern0pt#1}}^H$}%
  \ifdim\ht0=\ht2 #3\else #2\fi
  }
\newcommand*\rel@kern[1]{\kern#1\dimexpr\macc@kerna}
\newcommand*\widebar[1]{\@ifnextchar^{{\wide@bar{#1}{0}}}{\wide@bar{#1}{1}}}
\newcommand*\wide@bar[2]{\if@single{#1}{\wide@bar@{#1}{#2}{1}}{\wide@bar@{#1}{#2}{2}}}
\newcommand*\wide@bar@[3]{%
  \begingroup
  \def\mathaccent##1##2{%
    \if#32 \let\macc@nucleus\first@char \fi
    \setbox\z@\hbox{$\macc@style{\macc@nucleus}_{}$}%
    \setbox\tw@\hbox{$\macc@style{\macc@nucleus}{}_{}$}%
    \dimen@\wd\tw@
    \advance\dimen@-\wd\z@
    \divide\dimen@ 3
    \@tempdima\wd\tw@
    \advance\@tempdima-\scriptspace
    \divide\@tempdima 10
    \advance\dimen@-\@tempdima
    \ifdim\dimen@>\z@ \dimen@0pt\fi
    \rel@kern{0.6}\kern-\dimen@
    \if#31
      \overline{\rel@kern{-0.6}\kern\dimen@\macc@nucleus\rel@kern{0.4}\kern\dimen@}%
      \advance\dimen@0.4\dimexpr\macc@kerna
      \let\final@kern#2%
      \ifdim\dimen@<\z@ \let\final@kern1\fi
      \if\final@kern1 \kern-\dimen@\fi
    \else
      \overline{\rel@kern{-0.6}\kern\dimen@#1}%
    \fi
  }%
  \macc@depth\@ne
  \let\math@bgroup\@empty \let\math@egroup\macc@set@skewchar
  \mathsurround\z@ \frozen@everymath{\mathgroup\macc@group\relax}%
  \macc@set@skewchar\relax
  \let\mathaccentV\macc@nested@a
  \if#31
    \macc@nested@a\relax111{#1}%
  \else
    \def\gobble@till@marker##1\endmarker{}%
    \futurelet\first@char\gobble@till@marker#1\endmarker
    \ifcat\noexpand\first@char A\else
      \def\first@char{}%
    \fi
    \macc@nested@a\relax111{\first@char}%
  \fi
  \endgroup
}
\makeatother

\def\figdirOld{./}
\def\figdirNew{./}
\def\figdirSketch{./}

\def\cal{\mathcal}

\def\numA{0.4}
\def\numB{0.22}
\def\numC{0.155}

\begin{document}

\title{Solving nonequilibrium dynamical mean-field theory using matrix product states}

\author{F. Alexander Wolf}
\affiliation{
Theoretical Nanophysics,
Arnold Sommerfeld Center for Theoretical Physics,
LMU Munich,
Theresienstrasse 37,
80333 M\"unchen, Germany}
\author{Ian P. McCulloch}
\affiliation{
Centre for Engineered Quantum Systems,
School of Physical Sciences,
The University of Queensland,
Brisbane, Queensland 4072, Australia}
\author{Ulrich Schollw\"ock}
\affiliation{
Theoretical Nanophysics,
Arnold Sommerfeld Center for Theoretical Physics,
LMU Munich,
Theresienstrasse 37,
80333 M\"unchen, Germany}

\date{\today}

\begin{abstract}
We solve nonequilibrium dynamical mean-field theory (DMFT) 
using matrix product states (MPS).
This allows us to treat much larger bath sizes
and by that reach substantially longer times (factor $\sim$ 2 -- 3)
than with exact diagonalization.  
We show that the star geometry of the underlying impurity problem 
can have substantially better entanglement 
properties than the previously favored chain geometry. 
This has immense consequences 
for the efficiency of an MPS-based description of general impurity problems:
in the case of equilibrium DMFT,  
it leads to an orders-of-magnitude speedup.
We introduce an approximation for the
two-time hybridization function 
that uses time-translational invariance, which  
can be observed after a certain relaxation time after 
a quench to a time-independent Hamiltonian.
\end{abstract}


\maketitle

\section{Introduction}

The dynamical mean-field theory (DMFT)\cp{metzner89,georges92,georges96,kotliar06} 
is among the most successful methods
to study strongly correlated electron systems in higher dimensions.
DMFT maps a lattice model such as the Hubbard model 
onto an effective impurity model, 
which can be solved at considerably lower numerical cost.
The resulting approximation becomes exact in the limit of infinite dimensions,\cp{metzner89}
and is usually good for three dimensional systems. 
In the past years the nonequilbrium formulation of DMFT (NEQDMFT),\cp{schmidt02,freericks06,aoki14}
which generalizes DMFT to the Keldysh formalism, 
has become widely employed.

To advance DMFT in the nonequilibrium regime,
one still needs efficient methods to solve the real-time dynamics of the
effective underlying impurity model far from equilibrium. Impurity solvers
that have been used so far include real-time continuous-time
quantum Monte Carlo,\cp{eckstein09} which is numerically
exact, but restricted to short times
due to the \tit{phase} problem. Furthermore, there are strong-
\cp{eckstein10} and weak-coupling expansions,\cp{eckstein10i,amaricci12,tsuji13} which 
are restricted to certain parameter regimes,
and a formulation of NEQDMFT, that is able treat
the steady-state case efficiently.\cp{arrigoni13}
Recently,
a Hamiltonian-based impurity solver scheme has
been developed, which maps the DMFT impurity
model onto a single-impurity Anderson model (SIAM)
with a finite number of bath orbitals.\cp{gramsch13} This could be solved 
with exact diagonalization in all parameter regimes. 
While the representation of the DMFT bath with a SIAM can be 
made exact for small times, it requires 
an increasing number of bath orbitals to reach longer
times.\cp{gramsch13,balzer14i}
The exponential scaling of the Hilbert space dimension
as a function of the number of bath orbitals therefore 
prohibits to acquire the dynamics at long time scales.

Various approaches exist to overcome 
this limitation in the representation of the wave function.
These notably include (time-dependent) DMRG,\cp{schollwock05,schollwock11}
which is based on a matrix product state (MPS) representation, 
and tensor-network representations of many-fermion states.\cp{verstraete04,vidal07}  
Recently, the so-called multiconfiguration time-dependent
Hartree method\cp{balzer14} was applied to solve the Hamiltonian representation of DMFT.
In this paper, we study the application of MPS-based methods to it.

The paper is organized as follows.
In \secref{secNEQDMFT}, we briefly give the basic definitions of nonequilibrium DMFT.
Motivated by the fact that the mapping on a SIAM in NEQDMFT is 
simple if the SIAM is in the star geometry, while it is unsolved for the chain geometry,
in \secref{secEQ},
we compare the entanglement properties for the two cases. As these should not
depend on whether Hamiltonian parameters are time-dependent or not, we do this for the
equilibrium case. Unexpectedly, we find that the star geometry can have much better entanglement
properties than the chain geometry.
In \secref{secResultsNEQ}, we numerically solve the NEQDMFT
and analyze the computational resources needed to do so. 
In \secref{secApprox}, we propose a specific extrapolation of the hybridization function
that uses time-translational invariance, which is reestablished after a certain relaxation phase
after a quench to a time-independent Hamiltonian. 
In \secref{secCon}, we conclude the paper.

\section{Basics of nonequilbrium DMFT}
\label{secNEQDMFT}

We aim to describe the real-time evolution of a lattice quantum 
many-body system such as the single-band Hubbard model
\begin{align}
H\tl{Hub}(t) = -v(t) \sum_{ij\sigma} c_{i\sigma}^\dagger c_{j\sigma}
+
U(t)
\sum_{i}
(n_{i\uparrow}-\tfrac12)
(n_{i\downarrow}-\tfrac12),
\label{eqHubbard}
\end{align} 
where $c_{i\sigma}^\dagger$
($c_{i\sigma}$) creates (annihilates) an electron with spin $\sigma$ on site $i$ of the crystal
lattice, $n_{i\sigma}$ is the spin-resolved density, $v(t)$ is the hopping energy, 
and $U(t)$ is the local interaction energy.

The central task of nonequilibrium DMFT based on the Keldysh formalism~\cite{keldysh64} is to
compute the local contour-ordered Green's function
\begin{align}  \label{eqGcontour}
 G_\sigma(t,t')=- i \langle \mathcal{T}_{\mathcal C}c_{\sigma}(t)c\dag_{\s}(t')\rangle_{S_\mathrm{loc}},
\end{align}
of an effective single-site impurity model that approximates the lattice model \eqref{eqHubbard}. 
The time arguments of contour-ordered functions lie on the L-shaped Keldysh contour ${\cal C}$, 
and $\langle T_{\cal C} \ldots\rangle_{S_\mathrm{loc}} \equiv \mathrm{Tr} [T_{\cal C} 
e^{S_\mathrm{loc}} \ldots]/\mathrm{Tr} [T_{\cal C} e^{S_\mathrm{loc}}]$ 
denotes the contour-ordered expectation value.\cp{aoki14} 
For real-time arguments as studied in this paper, though, different orderings on the L-shaped contour 
simply lead to the familiar definitions of retarded and advanced Green functions. 
The action $S_\mathrm{loc}$ of the effective model is given by ($\hbar\equiv1$)
\begin{multline}
\label{eqDmftaction}
S_\text{loc} = -i \int_{\mathcal C} d t
\Big(U(t)(n_{\upa}(t)-\tfrac12)(n_{\dna}(t)-\tfrac12)-\mu\sum_\sigma n_\sigma(t)\Big)
\\
-i\int_{\mathcal C}
\int_{\mathcal C}
dt\,d t'\sum_{\sigma} c\dag_{\sigma}(t) \Lambda_{\sigma}(t,t') c_{\sigma}(t'),
\end{multline}
where the first part describes the local energies associated with the  
impurity ($\mu$ denotes the chemical potential), 
and the second part describes the hybridization of the impurity
with a bath of non-interacting fermions. 
This Gaussian bath is integrated out and by that gives rise to the two-time
hybridization function $\Lambda_\sigma(t,t')$. $\Lambda_\sigma(t,t')$ must be determined
self-consistently such that the resulting self-energy of the effective impurity model
equals the local self-energy of the lattice model. In the simplest case of a Bethe lattice
with nearest-neighbor hopping in the limit of infinite coordination number $Z$, 
this requirement leads to a self-consistency relation of closed
form\cite{eckstein09i}
\begin{equation}
\label{eqBethe}
\Lambda_\sigma(t,t') =  v(t) G_\sigma(t,t') v(t')\,,
\end{equation}
where the hopping matrix elements in \Eqref{eqHubbard} have been rescaled according to 
$v(t)\rightarrow v(t)/\sqrt{Z}$.\cp{metzner89} 

\subsection{Hamiltonian representation}

The DMFT action  $S_\mathrm{loc}$ in \Eqref{eqDmftaction} 
can also be represented by a time-dependent 
Anderson model (SIAM)\cp{gramsch13} 
\begin{align}
	\label{eqSiam}
	H(t) & = H_\mathrm{imp}(t)+H_\mathrm{bath}(t)+H_\mathrm{hyb}(t), \non\\
	H_\mathrm{imp}(t) & = U(t)\left(n_{0\uparrow}-\tfrac12\right)\left(n_{0\downarrow}-\tfrac12\right) -\mu \sum_\sigma n_{0\sigma},\nonumber\\
	H_\mathrm{bath}(t) & = \sum_{l=1}^{L_b}\sum_{\sigma} \epsilon_{l\sigma} c\dag_{l\sigma}c_{l\sigma},\nonumber\\
	H_\mathrm{hyb}(t) & = \sum_{l=1}^{L_b}\sum_{\sigma}\left(V_{l\s}(t)c\dag_{0\sigma}c_{l\sigma}+\text{H.c.}\right),
\end{align}
where the impurity at site $0$ is coupled 
with hopping energies $V_{l\s}(t)$
in a star geometry to $L_b$ 
noninteracting bath orbitals at potentials $\epsilon_{l\sigma}$,
which can be chosen to be time-independent.\cp{gramsch13}  
The hybridization function of a SIAM is
\begin{align}
\label{eqSIAMhyb}
 \Lambda\th{SIAM}_\sigma(t,t')=\sum_{l=1}^{L_b}V_{l\s}(t)g(\epsilon_{l\sigma},t,t')V_{l\s}(t')^*,
\end{align}
where 
\eq{ \label{eqGfree}
g(\epsilon_{l\s},t,t')=- i \big(\theta_{\cal C}(t,t')-f(\epsilon_{l\s})\big) \mathrm{e}^{-i\epsilon_{l\s}(t-t')}
}
is the Green function of an isolated bath orbital, $f(\epsilon)=1/(\mathrm{e}^{\beta\epsilon}+1)$ denotes 
the Fermi distribution, and $\theta_{\cal C}(t,t')$ is the contour step function
\eq{
\theta_C(t,t')=\left\{\begin{aligned}
	&1\text{ ~~for }t\geq_Ct'\\
	&0\text{ ~~else.}
\end{aligned}\right.
}

\subsection{How to obtain the Hamiltonian parameters?}
\label{secHamPars}

It remains to solve the following problem: 
Given the hybridization function $\Lambda_\sigma(t,t') =  v(t) G_\sigma(t,t') v(t')$, obtained from 
the self-consistency condition \eqref{eqBethe}, one needs to determine the  
Hamiltonian parameters of the SIAM \eqref{eqSiam} 
that generate this hybridization function 
$\Lambda\th{SIAM}_\sigma(t,t') = \Lambda_\sigma(t,t')$ via \Eqref{eqSIAMhyb}. 

To achieve this,\cp{eckstein09thesis,gramsch13} two distinct baths have to be introduced:
the \tit{first} bath $\Lambda_\s\th{SIAM,--}$ describes initial correlations in the system,
whereas the \tit{second} bath $\Lambda_\s\th{SIAM,+}$ describes the dynamic build-up of correlations.
The parameters $V_{l\s}(t)$ and $\e_{l\s}$ that generate the \tit{first} bath can be directly 
expressed using the bath spectral function that corresponds to $\Lambda(t,t')$. 
The parameters for the \tit{second} bath have to be constructed using a matrix factorization of $\Lambda(t,t')$.
As in this work, for simplicity, only time-evolutions from uncorrelated initial states
are considered, $\Lambda_\s\th{SIAM,--}(t,t')\equiv 0$, and we only recapitulate
the construction of  the \tit{second} bath $\Lambda_\s\th{SIAM,+} \equiv \Lambda_\s\th{SIAM}$.
 
In this case, it will be sufficient to consider Green's functions and the
hybridization function only for real-time arguments. For real times, 
rewriting the contour-ordered Green function \eqref{eqGcontour}
using the \tit{greater} and \tit{lesser}  Green function and introducing
an analogous definition for the hybridization function, leads to
\eqs{
G(t,t') & =  \theta_C(t,t') G^{>}_{\sigma}(t,t')+\theta_C(t',t)G^{<}_{\sigma}(t,t'),  \\
\Lambda_{\sigma}(t,t') & =  \theta_C(t,t')\Lambda^{>}_{\sigma}(t,t')+\theta_C(t',t)\Lambda^{<}_{\sigma}(t,t'), \label{eqLambdaSplit}
}
where
\eqs{  \label{eqGgtr}
  G^>(t,t')  & =  -i \langle c(t) c^\dagger(t') \rangle_{S\tl{loc}}, \\
  G^<(t,t') &   =   i \langle c^\dagger(t') c(t) \rangle_{S\tl{loc}}.
}       
This allows to rewrite the self-consistency \eqref{eqBethe} as
\eq{
\Lambda_\sigma^{\gtrless}(t,t') =  v(t) G_\sigma^{\gtrless}(t,t') v(t').
}

Independent of that, $\Lambda\th{SIAM}(t,t')$ in \eqref{eqSIAMhyb} can be simplified   
due to a freedom in choice for the bath potentials $\epsilon_{l\s}$,
which are chosen to have different initial and final constant values.\cp{gramsch13} 
Choosing $\epsilon_{l\s}=0$ for the final value cancels 
the oscillatory term $e^{-i\e_{l\s}(t-t')}$ in \Eqref{eqGfree}.
Considering occupied sites with initial potential energy $\epsilon_{l\s}<0$ at $T=0$, 
one has $g(\epsilon_{l\s},t,t')=-i(\theta_{\cal C}(t,t') - 1)=i\theta_{\cal C}(t',t)$,
whereas for unoccupied orbitals with initial $\epsilon_{l\s}>0$ 
one has $g(\epsilon_{l\s},t,t')=-i \theta_{\cal C}(t,t')$.

Rewriting \Eqref{eqSIAMhyb} with this choice\cp{gramsch13} 
for the potential energies gives 
\eq{
 \Lambda\th{SIAM}_\sigma(t,t') 
 = \,& i\theta_{\cal C}(t',t) \sum_{l \tx{ occ.}}V_{l\s}(t) V_{l\s}(t')^*  \non\\
&- i\theta_{\cal C}(t,t') \sum_{l \tx{ unocc.}}V_{l\s}(t) V_{l\s}(t')^*.
}
Comparison with \Eqref{eqLambdaSplit} then allows to rewrite the
self-consistency for the \tit{greater}
and \tit{lesser} hybridization functions as 
\begin{subequations}
\label{eqHybDecomp}
\eq{
 \Lambda^<_\sigma(t,t') & = i \sum_{l=1}^{L_b/2}V_{l\s}(t) V_{l\s}(t')^*, \\
 \Lambda^>_\sigma(t,t')  & = -i \sum_{l=L_b/2+1}^{L_b}V_{l\s}(t) V_{l\s}(t')^*,  
}
\end{subequations}
where we assumed the first half of bath orbitals to be occupied, 
and the second half to be unoccupied. If one can solve these
equations for the couplings $V_{l\s}(t)$, the  
construction of the appropriate SIAM is completed. 
In the limit $L_b\rightarrow\infty$ one can always find functions $V_{l\s}(t)$ that allow
to represent the two-time functions $\Lambda_\s^{\gtrless}(t,t')$ via \Eqref{eqHybDecomp}. 
For a finite number of bath sites, this is not guaranteed, and approximation methods have to be used. 
The method of choice\cp{gramsch13} is a Cholesky factorization of the matrices 
$\pm i\Lambda_\s^{\gtrless}(t,t')$ (the two-time function becomes a matrix with two discrete indices upon  
time discretization) combined with an optimization procedure.\cp{gramsch13}
Both are standard numerical routines and straight-forwardly give the 
Hamiltonian parameters $V_{l\s}(t)$.

\subsection{Equilibrium case}
\label{secIntroEQ}

To make the connection with exisiting treatments of 
DMFT calculations with DMRG,\cp{garcia04,nishimoto04,nishimoto04i,karski05,garcia07,karski08,ganahl14,wolf14,ganahl14i} 
we give the equations for the equilibrium case.
In equilibrium, 
all relevant two-time functions are time-translationally invariant and become
functions of effectively one time argument, \eg
\eq{ \label{eqGT0}
G_\s^R(t,t') & =  -i \theta(t-t') \big(G^>(t,t') - G^<(t,t')\big) \\
& \equiv G_\s^R(t-t') \non
}  
for the retarded component of the Green function.
One can therefore consider the corresponding 
one-argument Fourier (Laplace) transformed representation of such functions, \eg
$G_\s(\w) = \int dt\, e^{i\w t} G_\s^R(t,0)$, 
which is analytic in the upper half complex plane $\{\w \big\vert \tx{Re}(\w) > 0; \w \in\;\mathbb{C}\}$,
or  $\Lambda_\s(\w) = \int dt\, e^{i\w t} \Lambda_\s(t,0)$.
The analogous self-consistency condition to \Eqref{eqBethe} then is 
\eq{ \label{eqBetheEq}
\Lambda_\s(\w) = v^2 G_\s(\w).
}
The Fourier transform of the hybridization function 
of the SIAM \Eqref{eqSIAMhyb} is
\eq{ \label{eqSIAMeq}
 \Lambda\th{SIAM}_\sigma(\w)=\sum_{l=1}^{L_b} \frac{ \abs{ V_{l\s} }^2 }{\w - \epsilon_{l\s}},
}
with now time-independent hybridization couplings $V_{l\s}$. 

When solving the self-consistency condition \eqref{eqBetheEq}
with the help of a Fourier transform of $G_\s^R(t,0)$, 
one has to know $G_\s^R(t,0)$ at all times, in particular for $\abs{t} \rightarrow \infty$.
If $G_\s^R(t,0)$ decays quickly to zero, this poses no computational problem. If 
not, as in the interesting case close to phase transitions, 
very long times have to be computed, which is a hard problem due to entanglement 
growth in DMRG.\cp{wolf14} 
By contrast, the solution of the nonequilibrium self-consistency condition \eqref{eqBethe} 
does not \tit{a priori} require to compute very long times, as it does not invoke a Fourier transform.
Instead, one solves the self-consistency on the time domain starting at 
short times going successively to longer times. This makes it well suited for a DMRG treatment.

\section{Entanglement in the star vs. chain geometry in equilbrium}
\label{secEQ}

The DMFT impurity Hamiltonian is not a physical  
but an effecitve model for which the 
only requirement is that the bath hybridization function $\Lambda(t,t')$ 
fulfills a DMFT self-consistency condition. 
Apart from this, there is no constraint, and one is \eg  
free to choose the geometry of the impurity problem. To our knowledge,
up to now, for MPS/DMRG treatments of impurity problems,\cp{nishimoto04,nishimoto04i,raas04,karski05,karski08,ganahl14,wolf14,ganahl14i,nuss14} 
only the \tit{chain} geometry has been considered, which is
also used in NRG. This is due to the common belief that long-range interactions
make any treament with MPS very inefficient as then \tit{area laws} do not hold true any more.
As discussed in the following, the star geometry of an impurity problem can nevertheless
be highly suitable for an MPS treatment.
For this analysis, we consider different SIAMs in equilibrium, 
as the fundamental entanglement properties of the geometry should not 
depend on whether Hamiltonian parameters are time-dependent or not. 
In this section, therefore, Green and hybridization functions are time-translationally invariant.

\subsection{Star and chain geometry}

The Hamiltonians of the SIAM in the star and the chain geometry read as
\begin{subequations}
\label{eqHamGeo}
\eq{
H\th{star} & = H\tl{imp} + H\tl{bath} + H\tl{hyb},  \label{eqHstar} \\
& H\tl{imp} = U \left(n_{0\uparrow}-\tfrac12\right)\left(n_{0\downarrow}-\tfrac12\right),\\
& H\tl{bath} = \sum_{l=1}^{L_b}\sum_{\sigma} \epsilon_{l} c\dag_{l\sigma}c_{l\sigma},\\
& H\tl{hyb} = \sum_{l=1}^{L_b}\sum_{\sigma}\left(V_{l} c\dag_{0\sigma}c_{l\sigma}+\text{H.c.}\right), \label{eqHhyb} \\
H\th{chain} & = H\tl{imp} + H\tl{pot} + H\tl{kin}, \\
& H\tl{pot} = \sum_{l=1}^{L_b}\sum_{\sigma} \widetilde\epsilon_{l} c\dag_{l\sigma}c_{l\sigma}, \\
& H\tl{kin} = \sum_{l=0}^{L_b-1}\sum_{\sigma}\left(\widetilde V_{l} c\dag_{l+1,\sigma}c_{l\sigma}+ \text{H.c.}\right).
} 
\end{subequations}

\begin{figure}
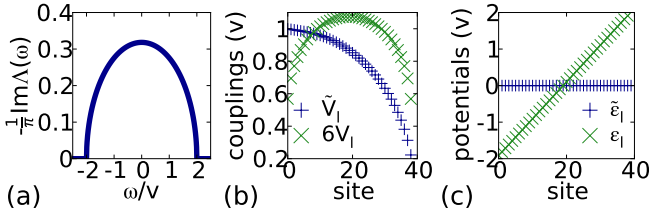

\ig{\numC}{pdf}{\figdirNew fig_SIAM_geo_Lambda}
\ig{\numC}{pdf}{\figdirNew fig_SIAM_geo_coupl}
\ig{\numC}{pdf}{\figdirNew fig_SIAM_geo_pots}
\caption{(Color online) 
Panel (a): Semi-elliptic bath spectral function 
$-\frac{1}{\pi} \tx{Im} \Lambda(\w)$ defined in \Eqref{eqSemiEll}.
Panel (b): Corresponding couplings in the chain ($\wt V_l$)
and the star ($V_l$) geometry.
Panel (c): Corresponding potentials in the chain ($\wt \e_l$)
and the star ($\e_l$) geometry.
Parameters are given in units of the hopping $v$.
}
\label{figSIAM_ChainStar_geo}
\end{figure}

$H\th{star}$ is a time-independent version of the representation 
of the SIAM chosen in the previous section in \Eqref{eqSiam}. 
The relation of both $H\th{star}$ and $H\th{chain}$ 
is a unitary transformation\cp{bulla08,raas05,wolf14}  
defined as the matrix of Lanzcos vectors that tridiagonalizes $H\th{star}$ 
(and hence maps it on a chain) as recapitulated in \appref{secMapStarChain}.

The hybridization functions of the SIAM in both geometries 
in their dependence on the Hamiltonian parameters of \Eqref{eqHamGeo} are,
\eq{ \label{eqSIAMgeos}
 \Lambda\th{star}(\w) & =\sum_{l=1}^{L_b} \frac{ \abs{ V_{l} }^2 }{\w - \epsilon_{l}}\\
 \Lambda\th{chain}(\w) & = \frac{\vert\widetilde V_0\vert^2}{\w - \wt\ve_1 - \displaystyle\frac{ \vert  \widetilde V_1 \vert^2}{\displaystyle \w- \wt\ve_2-\frac{\cdots}{\w-\wt\ve_{L_b-1} -\frac{\displaystyle \vert\widetilde V_{L_b-1}\vert^2}{\w-\wt\ve_{L_b}}}}}, \non
}
where the first line has already been given in \Eqref{eqSIAMeq}.

Consider now the example of a SIAM with a semi-elliptic bath spectral function,
which is given by the imaginary part of the hybridization function $\Lambda(\w+i0^+)$, where here $\w\in\mathbb{R}$,
\eq{  \label{eqSemiEll}
-\frac{1}{\pi} \tx{Im} \Lambda(\w+i0^+) = \frac{1}{2v\pi} \sqrt{4 - \Big(\frac{\w}{v}\Big)^2}
}
and shown in \figref{figSIAM_ChainStar_geo}(a). 
In the following, we will omit to specfiy the infinitesimal shift $i0^+$.
To find the parameters of the SIAMs that generate this 
hybridization function via \Eqref{eqSIAMgeos},  
one discretizes $-\frac{1}{\pi} \tx{Im} \Lambda(\w)$ in a 
procedure well known from NRG, which is 
briefly summarized in \appref{secBathDiscretization}.\cp{bulla08,wolf14}
The potentials $\e_{l}$ in the star can therefore be associated with excitations
of particles in different energy intervals of the bath spectral function $-\frac{1}{\pi} \tx{Im} \Lambda(\w)$, 
but have no simple interpretation in the chain geometry.
The resulting parameters are shown in \figref{figSIAM_ChainStar_geo}(b) and (c).

\begin{figure}
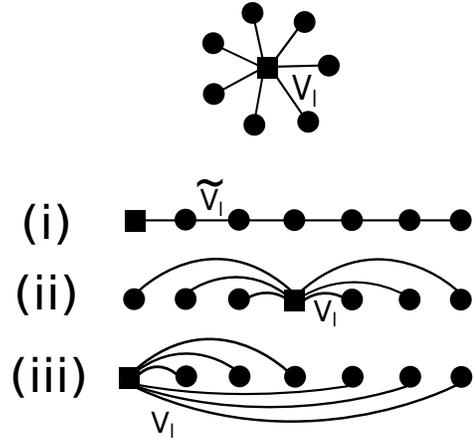

\ig{0.35}{pdf}{\figdirSketch geosNew}
\caption{(Color online)
Sketch of the three setups studied.
The star geometry can be mapped with the 
unitary transform $U$ to the \tit{chain} geometry (i).
It can also be mapped to an  \tit{auxiliary} chain
by sorting the indices ascendingly to their potential
energy. If one places the impurity in the center
of this chain, one obtains the layout (ii), if one
places it on the left edge of the chain, layout (iii)
is obtained.
Layouts (ii) and (iii) differ by the range over which the
couplings $V_l$ couple different lattice sites.
}
\label{figGeosSketch}
\end{figure}

We impose an order on the indices of the star bath states 
by sorting them according to their potential energy
in ascending order (\figref{figSIAM_ChainStar_geo}(c)),
which maps the star on an \tit{auxiliary chain} which should not be confused
with the \tit{chain geometry} introduced before. 
The decisive difference between the \tit{auxiliary chain} and the   
\tit{chain geometry} is that the former has long-range interactions
while the latter has short range interactions. We compare the case of the 
\tit{chain geometry} (i) 
with two different maps to generate the \tit{auxiliary chain}: 
(ii) placing the impurity site at the center, and (iii),   
placing the impurity at the first site. 
The \tit{auxiliary chain} obtained in case (iii) 
has long-range interactions at double the range of those that occur in case (ii).
One might expect this to lead to very different entanglement properties. 
All three cases are sketched in \figref{figGeosSketch}.
\begin{figure}
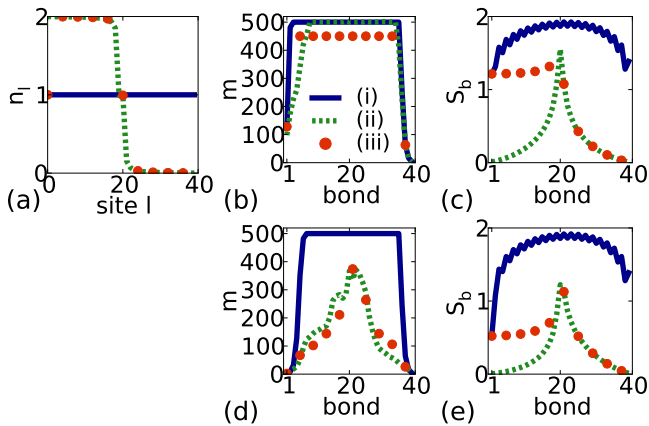

\ig{\numC}{pdf}{\figdirNew fig_SIAM_gs_dens}
\ig{\numC}{pdf}{\figdirNew fig_SIAM_gs_m}
\ig{\numC}{pdf}{\figdirNew fig_SIAM_gs_entropy}
\hspace*{\numC\textwidth}
\ig{\numC}{pdf}{\figdirNew fig_SIAM_init_m}
\ig{\numC}{pdf}{\figdirNew fig_SIAM_init_entropy}
\caption{(Color online) 
Panels (a) -- (c): Properties of the MPS approximation $\ket{E_0}$
of the ground state of a SIAM in the chain (full blue lines) 
in the three setups sketched in \figref{figGeosSketch}: 
(i) chain geometry (full blue lines), 
(ii) and (iii) star geometries with central and edge impurity 
(dashed green lines and red dotted lines).
This is for the semi-elliptic bath spectral function \eqref{eqSemiEll}
shown in \figref{figSIAM_ChainStar_geo}(a) and for $U/v=4$. 
Panel (d) and (e): Properties of the initial 
state $c_{0\s}\dag\ket{E_0}$ for the time evolution 
needed to compute the retarded Green function. 
Panel (a): Density distribution $n_l$.  
Panel (b) and (d): Bond dimension $m$. 
Panel (c) and (e): Bond entanglement entropy $S_b$.
$L_b=39$ sites are used to approximate the bath. 
The total chain length is $L=L_b+1=40$. 
Ground states have been computed with a maximum bond dimension 
of $m=500$. In the case of the chain geometry (i) this sufficed to reach a variance of 
$\expec{((H\th{chain}-E_0)/v)^2}\sim10^{-4}$, 
whereas in the case of the star geometry, (ii) and (iii), one could
reach $\expec{((H\th{star}-E_0)/v)^2}\sim10^{-6}$. 
Here, $E_0$ denotes the numerical value 
of the ground state energy.
}
\label{figSIAM_ChainStar_eq}
\end{figure}
\subsection{Ground state properties}

\figref{figSIAM_ChainStar_eq}(a)  shows the density distribution
in the ground state for the three setups (i) -- (iii). 
In the star geometry, \ie its \tit{auxiliary chain} representations (ii) and (iii),
the density distribution resembles the Fermi function,  
where sites with negative potential energy are occupied and 
sites with positive energies are unoccupied.
By contrast, the homogeneous potential energies of 
the chain geometry lead to a homogeneous density distribution. 
Whereas the wavefunctions of electrons,  which are non-interacting on all but one site of the system,   
is localized in the strongly inhomogeneous occupied regions in the star geometry,  
they are completely delocalized in the case of the chain geometry. 
Localization leads to low entanglement\cp{pekker14} and low bond dimensions,  
whereas delocalization leads to high entanglement. 
A similar observation can be made when comparing the momentum representation 
of free fermions, which is not entangled, with the real-space representation,
which is highly entangled.  
The fact that locality of the ground state in the star geometry 
transforms to non-locality in the chain geometry
is also obvious from inspection of the concrete unitary transform, which
is \tit{not} a Fourier transform, but still associates a superposition 
of all star bath states with a single chain bath state (see \eg \Eqref{eqFirstChainState} 
in \appref{secMapStarChain}).
\tit{Locality} is therefore not related to the range of interactions in this case.
We note that recent progress in 
exact diagonalization techniques also points out the fact that efficient
bath geometries should be designed in way that avoids partially filled
bath geometries.\cp{lu14}

\figsref{figSIAM_ChainStar_eq}(b) and (c) show the bond dimensions $m$ 
in the ground state and the bond entanglement entropies $S_b$.
These support the previous conceptual arguments when taking into account
that, in the case of the chain geometry (i) a maximum of $m=500$ kept states 
sufficed to reach a variance of 
$\expec{((H\th{chain}-E_0)/v)^2}\sim10^{-4}$, 
whereas in the case of the star geometry, (ii) and (iii), one could
reach the much better value of $\expec{((H\th{star}-E_0)/v)^2}\sim10^{-6}$. 

\subsection{Time evolution}

To understand how entanglement grows
during time evolution, 
consider the computation of the \tit{greater} 
Green function for the impurity (compare its definition \Eqref{eqGgtr})
\eq{
G_{\s}^{>}(t,t') = -i\expec{c_{0\s}(t) c_{0\s}\dag(t')}
} 
where the expectation value at $T=0$ is taken in the ground state.
In equilibrium, where $G_{\s}^>(t,t') = G_{\s}^>(t-t')$, 
one can without loss of generality set $t'=0$ and instead compute 
\eq{ \label{eqGgtrEvol}
G_{\s}^{>}(t) = -i\, \bra{E_0} c_{0\s}\,e^{-i(H-E_0)t} c_{0\s}\dag \ket{E_0}.
} 

\subsubsection{Initial state}

Applying the creation operator $c_{0\s}\dag$ to the ground state
destroys much of its entanglement, as can be seen
by inspecting \figref{figSIAM_ChainStar_eq}(d) and (e), which 
show bond dimensions and entanglement in the initial state $c_{0\s}\dag \ket{E_0}$ used
for the time evolution in \Eqref{eqGgtrEvol}. 
The action of $c_{0\s}\dag$ on the ground state $\ket{E_0}$
cancels \tit{exactly} all superpositions of Fock states in which the 
impurity site is occupied, which strongly reduces entanglement.
As the impurity site is involved in almost all states in the star geometry, 
the action $c_{0\s}\dag$ reduces entanglement in the star geometry dramatically, 
almost independently of whether the site is located  
at the center (ii) or at the edge (iii) (\figref{figSIAM_ChainStar_eq}(d)). 
In the chain geometry, the site does not have such a prominent 
role and therefore, reduction of entanglement is much less 
pronounced (\figref{figSIAM_ChainStar_eq}(d)).
\begin{figure}
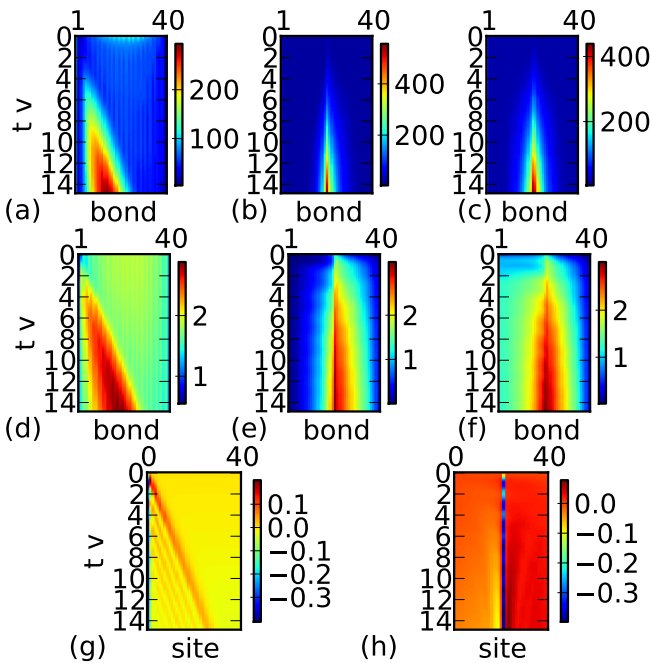

\ig{0.48}{pdf}{\figdirNew fig_SIAM_evol_m}
\ig{0.48}{pdf}{\figdirNew fig_SIAM_evol_entropy}
\ig{0.48}{pdf}{\figdirNew fig_SIAM_evol_dens}
\caption{(Color online) 
Time evolution of a SIAM with semi-elliptic 
bath spectral function \eqref{eqSemiEll} at $U/v=4$ for the
three setups (i) -- (iii) shown in \figref{figGeosSketch}.  
The properties of the initial states for the evolution are shown 
in \figref{figSIAM_ChainStar_eq}. 
Panels (a), (d) and (g) 
refer to the chain geometry (i), 
panels (b), (e) and (h) refer to the star geometry 
with the impurity located at the center (ii), 
and panels (c), (f) to the star geometry
with the impurity located at the left edge (iii). 
Panels (a), (b) and (c)  show the local 
bond dimension $m$ plotted versus bond and time,
panels (d), (e) and (f) show 
the bond entanglement entropy $S_b$, and panels (g) and (h) show 
the density distribution substracted from its initial value $n_l(t)-n_l(0)$
plotted versus site $l$ and time.
}
\label{figSIAM_chainStar_neq}
\end{figure}
\begin{figure}
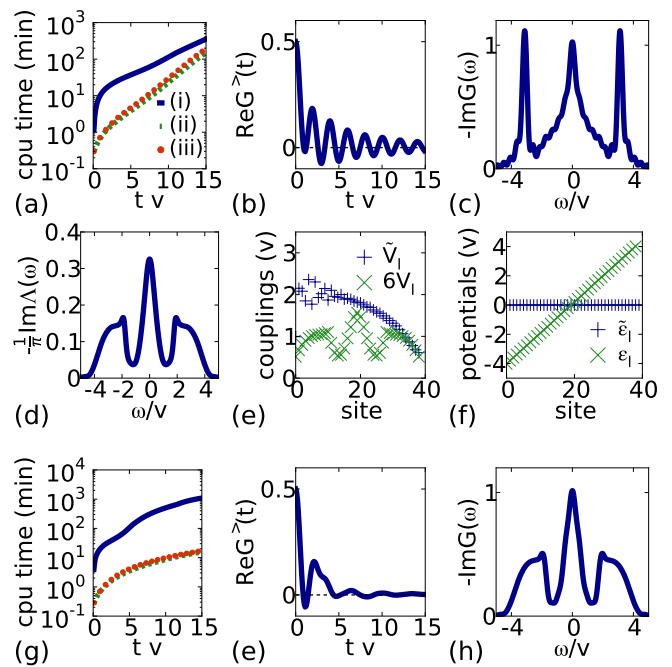

\ig{\numC}{pdf}{\figdirNew fig_SIAM_evol_cput}
\ig{\numC}{pdf}{\figdirNew fig_SIAM_evol_G}
\ig{\numC}{pdf}{\figdirNew fig_SIAM_evol_Gfour}
\vspace{1em}
\ig{\numC}{pdf}{\figdirNew fig_DMFT_geo_Lambda}
\ig{\numC}{pdf}{\figdirNew fig_DMFT_geo_coupl}
\ig{\numC}{pdf}{\figdirNew fig_DMFT_geo_pots}
\ig{\numC}{pdf}{\figdirNew fig_DMFT_evol_cput}
\ig{\numC}{pdf}{\figdirNew fig_DMFT_evol_G}
\ig{\numC}{pdf}{\figdirNew fig_DMFT_evol_Gfour}
\caption{(Color online) 
Panels (a) -- (c) refer to the computation with
semi-ellipitic bath spectral function (\figref{figSIAM_ChainStar_geo}(a)). 
Panels (d) -- (h) refer to a computation with the self-consistently 
determined bath spectral function (panel (d)) for $U/v=4$.
Panels (a) and (g) show computation time versus physical time. 
Therein, the blue solid line refers to the chain geometry (i) whereas  
the green (red) dashed (dotted) line refers to the 
star geometry with the impurity located at the center (ii) 
(at the edge (iii)).
Panels (b) and (h) show the time evolution of the greater
Green function and panels (c) and (i) show the 
Fourier transform of the Green function,
which is the same for all three setups (i) --- (iii),
and therefore only one curve is shown.
The oscillations in the resolution of $-\tx{Im}G(\w)$ in panel (c)
can be removed by convolution with a Gaussian or a Lorentzian of small
width $\eta$. On the (real-) time domain,
this would correspond to a slight damping (\tit{windowing}) of 
$G^>(t)$ with a Gaussian or Lorentzian 
of large width $1/\eta$ and maximum at $t=0$.
This suppresses contributions for times $t\gtrsim 1/\eta$. 
Alternatively, one can 
compute the real time evolution of the 
Green function up to higher times,
until it has converged to zero, or use an extrapolation 
technique such as \tit{linear prediction}.\cp{numrec07,white08,barthel09,ganahl14i} 
}
\label{figSIAM_chainStar_neq2}
\end{figure}
\subsubsection{Entanglement growth}
\label{secEntGrowth}

During the real time evolution needed  
to compute \Eqref{eqGgtrEvol}, 
we compute each time step $\D t = 0.05/v$ with a precision of 
\eq{ \label{eqErr}
\e\tl{err} = \vert\vert\,\ket{\psi(t+\D t)} - \exp(-i H \D t) \ket{\psi(t)} \,\vert\vert < 10^{-6},
}
and do not limit the growth of bond dimensions 
needed to guarantee this error. 
Truncating the initial state down to this precision 
reduces the original bond dimensions shown 
in  \figref{figSIAM_ChainStar_eq}(d)  
to very small values. These can be seen in the short-time regions   
of \figsref{figSIAM_chainStar_neq} (a) -- (c),
where we plot the bond dimensions that occur 
in the three setups (i), (ii) and (iii), respectively.

In the chain geometry, the growth of the bond dimension $m$ 
(\figref{figSIAM_chainStar_neq} (a)) and of the entanglement 
entropy $S_b$ (\figref{figSIAM_chainStar_neq} (d)) 
is associated with the particle that is created at
site 0 at time $t=0$ and subsequently travels across the chain as
seen by its density evolution shown in \figref{figSIAM_chainStar_neq}(g). 
In the regions that have not yet been reached 
by the particle, almost no change in $m$ and $S_b$ is observed. 
In the star geometry, by contrast, the particle remains almost 
localized (\figref{figSIAM_chainStar_neq}(h)) 
and entropy grows much more locally (\figsref{figSIAM_chainStar_neq}(e) and (f)).  
$m_b$ and $S_b$ are peaked at the center of the system 
as entanglement builds up only with 
low-energy states during time-evolution. 
These low-energy bath states are located at the center of
the system irrespective of whether the impurity 
is located there (ii) or at the edge (iii). The build-up of entanglement with high-energy
bath states would involve the occupation of these states, which is energetically
strongly suppressed.
\begin{figure}[t]
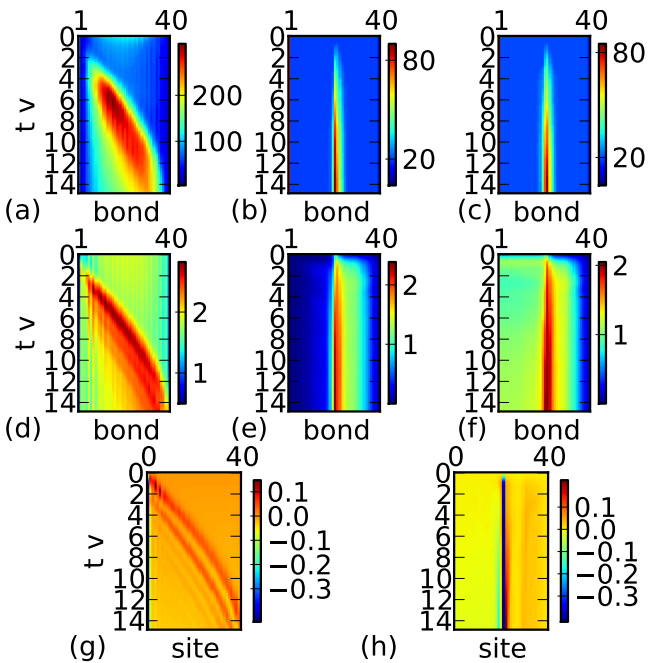

\ig{0.48}{pdf}{\figdirNew fig_DMFT_evol_m}
\ig{0.48}{pdf}{\figdirNew fig_DMFT_evol_entropy}
\ig{0.48}{pdf}{\figdirNew fig_DMFT_evol_dens}
\caption{(Color online) 
Time evolution for SIAMs in both geometries for a SIAM 
with self-consistently determined bath spectral function as shown in \figref{figSIAM_chainStar_neq2}(d).
The definition of panels is analogous to the one of \figref{figSIAM_chainStar_neq}.}
\label{figDMFT_chainStar_neq}
\end{figure}

\figref{figSIAM_chainStar_neq2}(a) then shows how this affects the computer time needed to reach a certain physical time.  
The chain geometry (i) is clearly less efficient than the star geometry setups (ii) and (iii).
\figref{figSIAM_chainStar_neq2}(b) and (c) show the time-evolution of the Green function
and its Fourier transform, which are identical in all three setups (i) --- (iii). 
All of the preceding results are not specific for 
the SIAM with semi-elliptic bath spectral function at $U/v=4$. In all other
cases studied by us, they are valid to an even greater extent. 
Consider the case with a bath spectral function $-\frac{1}{\pi} \tx{Im}\Lambda(\w)$ that 
is the solution of the DMFT for the Bethe lattice for $U/v=4$ as shown in   
\figref{figSIAM_chainStar_neq2}(d). The qualitative form of the Hamiltonian parameters,
shown in \figsref{figSIAM_chainStar_neq2}(e) and (f), is 
still similar to the previous case (\figref{figSIAM_ChainStar_geo}), but 
the absolute magnitude of couplings and potentials 
is higher as the support of the bath spectral function 
is now of the order of $2 U = 8 v$. 
This makes the system in the star geometry more 
inhomogeneous and thus more localized
--- entanglement between sites of very different energy is disfavored energetically --- whereas no 
such effect occurs in the chain geometry. 
Therefore, one sees that all computations can be performed
with tremendously increased efficiency in the star geometry, which results in 
computation time reductions of \tit{two orders of magnitude} as shown 
in \figref{figSIAM_chainStar_neq2}(g).
\figref{figDMFT_chainStar_neq}, which is organized in the same way as \figref{figSIAM_chainStar_neq},
shows that this speedup comes with much lower bond dimensions than
in the previous case (\figref{figSIAM_chainStar_neq}).

It remains to consider different values for the interaction $U$. 
It turns out that the intermediate value $U/v=4$ 
leads to the strongest entanglement growth. 
For low and high values of the interaction $U$, all preceding arguments still hold true,
but entanglement growth is strongly reduced in the star geometry 
for two further important reasons.
In the non-interacting limit, $U=0$, 
$c_{0\s}\dag\ket{E_0}$ is an eigentstate of $H$. 
This implies that time evolution does not affect entanglement in the state.
By continuity, \tit{close} to the 
non-interacting limit, only very few entanglement
is generated. In the strongly-interacting limit $U\gg v$,
very low entanglement growth is observed for a different reason.
Excitation of high energy states, \ie occupation of bath 
sites with high energy in the star geometry, 
has to involve hopping across the impurity, where the electron
needs to pay the energy for double occupation $U$. For high 
values of $U$, this process is strongly suppressed, and the consequence
is again much lower bond dimensions than in the intermediate case $U/v=4$.
In \appref{secGeosInt}, these arguments are supported with 
numerical data for the bond dimensions 
(\figref{figGeosIntm}) for all interaction strengths.

\subsection{Nature of long-range interactions}
\label{secLongRange}

The preceding arguments and observations show 
that the long-range interactions
present in the \tit{auxiliary chain} representation 
of the \tit{star} geometry 
do not imply that it is \tit{a priori} less suited for the 
treatment with MPS 
than the short-range interacting \tit{chain} geometry. 
One should realize that the long-range interactions
in the \tit{auxiliary chain} are \tit{not physical} interactions as they do \tit{not} 
occur among \tit{all} sites separated by a certain interaction range. 
They are \tit{artificial} interactions 
that occur \tit{exclusively} between the impurity site and 
each single bath site.
If this were not the case, the calculations using the 
\tit{auxiliary chain} with the impurity at the center (ii) 
and the left edge (iii) of the system should lead to very different entanglement, 
as the second case has long-range interactions 
at double the distance than in the first case. 
But as obvious from all examples discussed before 
(see \eg the plots for the computer time in \figref{figSIAM_chainStar_neq2}(a) and (g)), 
entanglement is comparable in both setups. 

The physical interpretation of the concept of entanglement entropy 
for these long-range interacting systems 
is no longer meaningful. There is no physical content in the 
notions \tit{left subsystem} and \tit{right subsystem} as 
in the usual line of argumentation 
when introducing DMRG, for instance, for the case of a Heisenberg spin chain. 
Still MPS can be a meaningful representation, 
but should then simply be interpreted as a certain way to 
manage and store the coefficients of the superpositions
of Fock states $\ket{\alpha_0 \alpha_1 \alpha_2 \dots}$ 
where $\alpha_i \in \{0,\uparrow,\downarrow,\uparrow\downarrow\}$ 
denotes the local quantum state. 
The corresponding MPS realizes, by computing all contractions of matrices
over physical quantum numbers $\{\alpha_i\}$ 
the subset of all possible $4^L$ Fock states, 
whose members have significant weight in a given 
many-body state $\ket{\psi}$.
Independent of whether the underlying Hamiltonian has long-range
interactions or not, bond dimensions in a given MPS can
be strongly reduced by reducing the number of Fock states with significant weights 
in $\ket{\psi}$. 
In the case of the strongly inhomogeneous 
problem of the star geometry, states that involve 
occupied sites with high potential
and unoccupied sites with low potential have a very small weight.
In the case of the homogeneous chain geometry,
no such argument applies, and \tit{a priori} the number of Fock 
states with significant weight can be much higher.

\section{Solving nonequilibrium DMFT using MPS}
\label{secResultsNEQ}

Having motivated the usage of the star geometry of impurity problems
for MPS based algorithms in the previous section, 
we will now use it to solve NEQDMFT. 
This point is important as a formulation of NEQDMFT in the chain geometry 
is highly non-trivial and has not yet been achieved,
whereas its formulation in the star geometry 
has been worked out by \ct{eckstein09thesis} and \ct{gramsch13}.

\subsection{Model definition}

In the following, we briefly summarize the benchmark setups 
studied by \ct{gramsch13} and \ct{balzer14} by means of exact
diagonalization and multiconfiguration time-dependent Hartree. 
Consider the NEQDMFT for the Hubbard model on the Bethe lattice, \ie impose  
the self-consistency condition \eqref{eqBethe}, for an initial preparation of the 
system in the atomic limit ($v$ $=$ $0$). 
The following ramp then rapidly turns on the hopping up to a 
final value of $v$ $=$ $v_0$ $\equiv$ $1$ at time $t_{1}>0$ 
\begin{align}
\label{eqRamp}
v(t)=\begin{cases}
		\tfrac{1}{2}(1-\cos(\omega_0 t))&\text{for~}t<t_{1},~\omega_0=\tfrac{\pi}{t_{1}}\\
		1&\text{for~}t\ge t_{1}.
\end{cases}
\end{align}
Since we start from the atomic limit, there are no impurity-bath correlations 
in the initial state and we only need to consider the 
\tit{second} bath, as discussed in \secref{secNEQDMFT}. 

The hybridization function $\Lambda(t,t')$ is particle-hole symmetric and 
spin-symmetric ($\Lambda_{\uparrow}=\Lambda_{\downarrow}=\Lambda$) 
in the para-magnetic phase considered here.  
The initial ground state of the SIAM contains an equal number of empty 
and doubly-occupied bath sites and a singly-occupied impurity. 
In practice, we average over two Green functions $G^\alpha$ and $G^\beta$, where the impurity of system 
$\alpha$ ($\beta$) is populated initially by a single up-spin (down-spin) electron.
The full Green function is then given by
\begin{align}
 G_{\sigma}(t,t')=\tfrac{1}{2}(G^\alpha_{0\sigma}(t,t')+G^\beta_{0\sigma}(t,t')).
\end{align}
Taking the average restores particle-hole symmetry, which is not given 
for $G^{\alpha}$ or $G^{\beta}$ alone. 

The self-consistency condition \eqref{eqBethe} is solved in the formulation \eqref{eqHybDecomp} 
by a matrix decomposition of $-i\Lambda^{<}(t,t')$ into coupling parameters $V_{l\s}(t)$, 
as explained in detail by \ct{gramsch13}.
Knowing the coupling parameters, we compute the real-time impurity Green functions 
$G^s_{\sigma}(t,t')=\theta_C(t,t')G^{s,>}_{\sigma}(t,t')+\theta_C(t',t)G^{s,<}_{\sigma}(t,t')$ 
with respect to the SIAMs $s=\alpha$ and $s=\beta$ by an MPS Krylov time evolution algorithm
\begin{align}
 G^{s,>}_{\sigma}(t,t')&=-i\bra{\psi_0^s}U(0,t)c_{0\sigma}U(t,t')c^\dagger_{0\sigma}U(t',0)\ket{\psi_0^s},&\nonumber\\
G^{s,<}_{\sigma}(t,t')&=i\bra{\psi_0^s}U(0,t')c^\dagger_{0\sigma}U(t',t)c_{0\sigma}U(t,0)\ket{\psi_0^s},\nonumber\\
U(t,t')&=\mathcal{T}_t \exp\Big(-i\int_{t'}^{t}\mathrm{d}s\,H(s)\Big),
\end{align}
where $\mathcal{T}_t$ denotes the usual time-ordering operator. 
For this, we use a simple middle-point approximation to evolve $\ket{\psi}$ one time step $\Delta t$ further, 
\eq{
\ket{\psi(t+\Delta t)} = \exp\big(-iH(t+\Delta t/2) \Delta t\big) \ket{\psi(t)}
}
and interpolate the Hamiltonian, \ie the couplings $V_{l\s}(t)$, with standard spline interpolation.

We compute the system's kinetic energy as
$E_\text{kin}(t)=-i\sum_\sigma\int\limits_C ds\Lambda(t,s) G_\sigma(s,t')|^<_{t=t'}$,
the density $\langle n(t)\rangle=-i\sum_\sigma G^<_\sigma(t,t)$, which is a conserved quantity, and 
the double occupation $d (t) =\langle n_{0\uparrow}(t)n_{0\downarrow}(t)\rangle$.
All of these quantities are averaged over the SIAMs $\alpha$ and $\beta$. The double occupation 
also gives access to the interaction energy, 
$E_\text{int} (t)=U( d(t) -\tfrac{1}{4})$ and by that
allows to compute the total energy as $E\tl{tot} = E\tl{kin}+E\tl{int}$.

In \secref{secHamPars}, we explained that we choose   
the bath potentials to be homogeneous. By this, a substantial part of the discussion
of \secref{secEQ} that was based on the inhomogeneity 
of the star geometry does not apply to the description of the present setup.
There are three arguments, that still motivate the use of the star geometry.
(a) The statements about the nature of long-range interactions interactions in \secref{secLongRange}
remain still valid and are independent of whether the problem is homogeneous or not. 
(b) If one does not start from the atomic limit, but has to consider initial correlations in the bath, 
this will require the representation of the 
\tit{first} bath referred to in \secref{secNEQDMFT}. This will again be inhomogeneous,
and all of the results of  \secref{secEQ} will again apply.
(c) The formulation of the nonequilibrium problem in the chain geometry is highly non-trivial, whereas
in the star geometry, computations can be carried out straight-forwardly.

\subsection{Numerical results}
\begin{figure}
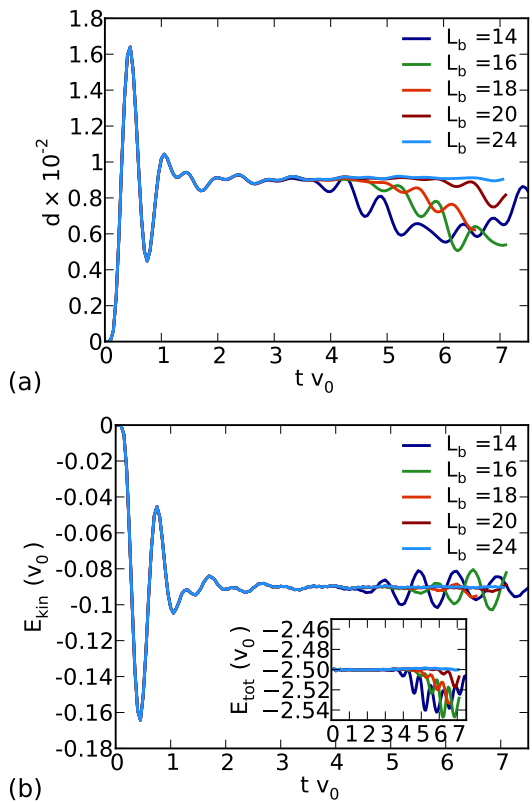

\ig{\numA}{pdf}{\figdirOld fig01_dU10}
\ig{\numA}{pdf}{\figdirOld fig02_EkinEtotU10}
\caption{(Color online) 
Time evolution of double occupancy $\expec{d(t)}$ (panel (a)) and kinetic and total energy (panel (b))
for $U=10$ and different bath sizes. For the largest bath shown $L_b=24$ we could 
reach a time $t\tl{max}\sim7/v_0$ in a controlled way, 
meaning that the total energy is conserved.
We did \tit{not} limit the maximal allowed bond dimension $m$, but 
we imposed an upper error bound for the time evolution
in single timestep of $\D t=0.05/v_0$ of $\e\tl{err}=10^{-6}$, as defined in \Eqref{eqErr}.
}
\label{figNEQDMFT_U10}
\end{figure}

\begin{figure}
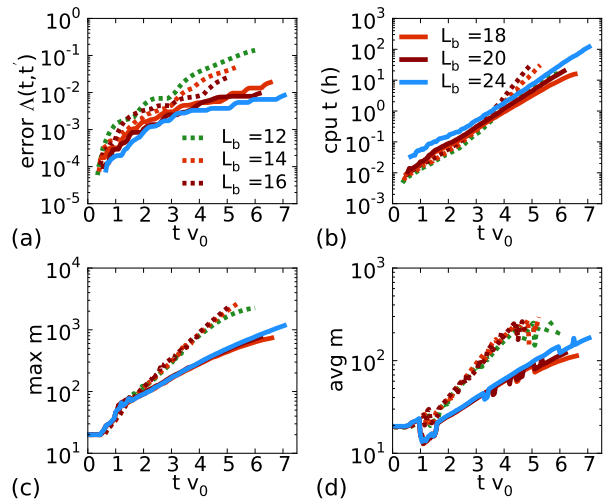

\ig{\numB}{pdf}{\figdirOld fig05_ErrLambda}
\ig{\numB}{pdf}{\figdirOld fig04_cput}
\ig{\numB}{pdf}{\figdirOld fig07_MaxM}
\ig{\numB}{pdf}{\figdirOld fig08_AvgM}
\caption{(Color online) 
Solid lines refer to the $U=10$ case, dashed lines to the $U=4$ case. 
Panel (a) shows the error 
$\max\abs{\Lambda(t,t')-\Lambda\tl{cholesky}(t,t')}$ 
of the hybridization function.
Panel (b) shows the computation time needed to compute one 
time step in the DMFT scheme of $\D t = 0.05/v_0$.
Panel (c) shows the maximal bond dimension that occurs
in set of Krylov states needed to expand $\exp(-iH(t)\Delta t)$.  
Panel (d) shows the average bond dimension
in these states. The average dimension is much lower
than the maximal dimension, which can be understood 
when looking at the spatially resolved
bond dimensions that are shown in \figsref{figSIAM_chainStar_neq} 
and \ref{figDMFT_chainStar_neq} for the equilibrium case,
but are typical also for the nonequilibrium case.
}
\label{figNEQDMFT_tech}
\end{figure}

\figref{figNEQDMFT_U10}(a) shows the time evolution of the double occupation
$d(t)$ for an interaction energy of $U=10$.
Whereas in exact diagonalization, the maximal treatable bath size was 
$L_b=14$,\cp{balzer14} we are able to perform computations
for $L_b=24$ in a numerically controlled way. 
The error measure for this is the conservation of the 
total energy $E\tl{tot}(t)$ shown in the inset of \figref{figNEQDMFT_U10}(b).
The bath size of $L_b=24$ allows to reach $t\tl{max}\sim 7/v_0$,
whereas the highest reached time in the literature up to now, for the case $U/v_0=10$, 
is $t\tl{max}\sim 2.5/v_0$.\cp{gramsch13} 
The substantial increase of the possible simulation time
is related to the reduced approximation error for the hybridization 
function $\max\abs{\Lambda(t,t')-\Lambda\tl{cholesky}(t,t')}$ 
for large bath sizes as shown in \figref{figNEQDMFT_tech}(a).  

In \figref{figNEQDMFT_tech}(b), we show the computer 
time needed to converge one DMFT time slice $\D t = 0.05/v_0$
using four DMFT iterations on a slice. The computation uses 
two cores, one for each SIAM $s=\{\alpha,\beta\}$.
\figref{figNEQDMFT_tech}(c) shows the maximal bond dimension
that occurs in the computed states to be around $m\sim 1000$ for the largest bath in the case 
of $U/v_0=10$. The average bond dimension, shown in \figref{figNEQDMFT_tech}(d),
is much lower, as the distribution of $m$ is strongly inhomogeneous, 
similarly to the cases studied before, see \eg \figref{figDMFT_chainStar_neq}(b).
The storage of MPS with these bond dimensions is easily feasable. 
The exponentially growing computation time in \figref{figNEQDMFT_tech}(b)
limits the accessible time scales. The shown accessed times though can still 
be reached comparatively easily, when realizing that computations on the $t$ - $t'$ - grid 
can be trivially parallelized with a linear speedup (computations
on one time slice are independent from each other). 
In practice, we used 16 cores to compute the time evolution for $L_b=24$ and $U=10$. 
All of the above used the $U(1)\times U(1)$ symmetry of the underlying SIAMs
that are associated with particle number conservation and the $S_z$ total spin. 
A computation that uses the $U(1)\times SU(2)$ symmetry should strongly
increase the computational efficiency. The maximally reachable simulation
time should then be around $t\tl{max} \sim 8/v_0 - 9/v_0$.

\begin{figure}
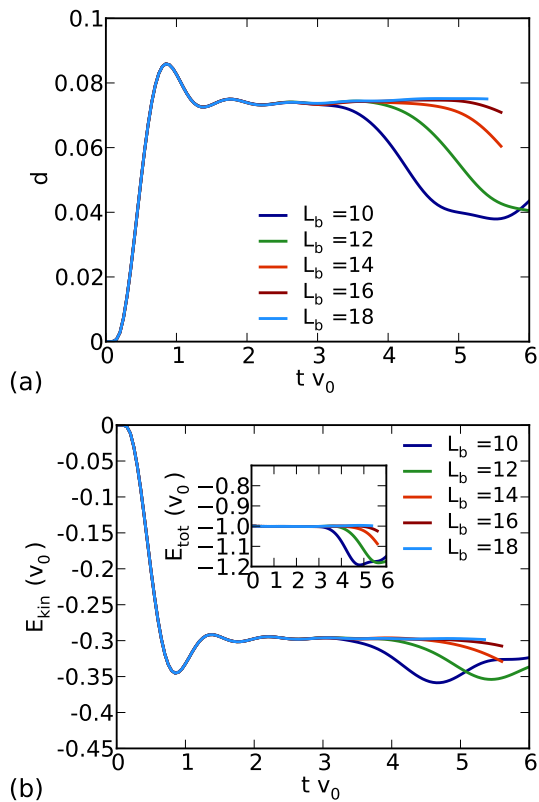

\ig{\numA}{pdf}{\figdirOld fig01_dU4}
\ig{\numA}{pdf}{\figdirOld fig02_EkinEtotU4}
\caption{(Color online) 
Time evolution of double occupancy $\expec{d(t)}$ (panel (a)) and kinetic and total energy (panel (b))
for $U=4$ and different bath sizes. For the largest bath shown $L_b=18$ we could 
reach a time $t\tl{max}\sim5.5/v_0$ in a controlled way, meaning that the total energy is conserved.
}
\label{figNEQDMFT_U4}
\end{figure}

Let us now study the much harder  
case of intermediate interaction strength $U/v_0=4$. 
\figref{figNEQDMFT_U4} shows results for the double occupation $d(t)$
and the kinetic and total energies for this case. 
Entanglement entropy grows much faster than for 
$U/v_0=10$, as mixing between occupied and empty bath orbitals
is energetically less suppressed, as discussed in \secref{secEntGrowth}.
This is reflected in the rapid growth of bond dimensions shown by the dashed lines
in \figref{figNEQDMFT_tech}(c) and (d).
Still the bath sizes treated here are beyond the regime
of exact diagonalization and multiconfiguration time-dependent hartree.\cp{balzer14}
While \ct{gramsch13} could reach $t\tl{max} \sim 2.8/v_0$ using exact diagonalization,
we reach $t\tl{max} \sim 5.5/v_0$ by investing the computational resources 
shown in \figref{figNEQDMFT_tech}. Also here, using higher computational resources,
and extrapolating the maximal bond dimension shown in \figref{figNEQDMFT_tech}(c)
would allow to reach $t\tl{max} \sim 6/v_0$ with a maximal $m\sim 10^4$. 
Again, usage of the SU(2) symmetry can help to substantially increase
the computational efficiency and increase the value of $t\tl{max} \sim 6/v_0$.

In \appref{secSimpleImp}, we study a simple time-dependent 
impurity problem as done by \ct{balzer14}, 
for which neither a selfconsistency DMFT loop has to be iterated nor does time
propagation need to be computed in the whole $t$ -- $t'$ plane. 
This reduces computation times by orders of magnitudes, 
and we could reach higher values of $t\tl{max}$. 
The entanglement growth observed for this example was comparable 
to the full self-consistent calculation.
In \appref{secGauge},
we study an inhomogeneous reformulation of the impurity problem,
motivated by the results of \secref{secEntGrowth} that showed that strongly
inhomogeneous impurity models lead to less entanglement
than homogeneous models. This reformulation can  
be easily achieved by using the local gauge symmetry of the couplings $V_{l\s}(t)$,
that has already been used to render the bath potentials time-independent
(see \secref{secHamPars}). We found, though, that the increased driving of the system 
that is implied by this reformulation exactly compensates the positive effect
of the inhomogenity, and by that, the same entanglement growth is observed 
in both setups.

\section{Approximated self-consistency: relaxation phase and steady phase}
\label{secApprox}

For a quench to a Hamiltonian that becomes time-independent 
after a certain transition period, 
one observes that, 
after a \tit{relaxation phase} that lasts until $t\tl{relax}$, 
the Green function shows the 
time-translational invariance that it would fulfill in 
equilibrium $G(t,t') = G(t-t')$. Putting that differently, 
it fulfills the symmetry $G(t,t') = G(t+s,t'+s)$ for some
intermediate time $s$ if $\min(t,t') > t\tl{relax}$, \ie 
$G(t+s,t'+s) = \const(t,t')$ can be extrapolated 
using a constant value.
By virtue of the self-consistency condition \eqref{eqBethe} 
the same argumentation holds true for the hybridization function,
and one can conlude that, as soon as time-translational invariance
is restored, one does no longer need to solve the DMFT self-consistency
on the whole time slice, but already \tit{knows} the correct $\Lambda(t,t')$
by extrapolation for times  $t' > t\tl{relax}$.
The DMFT iteration needs only to be computed
for  ``small" times on the time slice $t'\in [0,t\tl{relax}]$,
whereas usually, one has to compute it for $t' \in [0,t]$. 

\begin{figure}
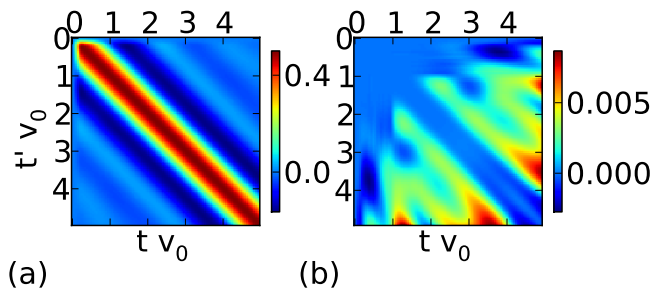

\ig{0.48}{pdf}{\figdirOld fig_offset_ILambda}
\caption{(Color online) 
Panel (a): hybridization function $\Lambda(t,t')$ obtained from a calculation in which the
self-consistency has been computed on the ``full" $t$--$t'$ grid. Panel (b):
Difference of hybridization functions $\Lambda\tl{relax}(t,t') - \Lambda(t,t')$,
where for $\Lambda\tl{relax}(t,t')$, the self-consistency has only been solved
for times $t'<t\tl{relax}=1/v_0$.}
\label{figRelaxLambda}
\end{figure}

\figref{figRelaxLambda}(a) shows a typical self-consistently
determined hybridization function $i\Lambda^{>}(t,t')$ 
for the same setup as studied in the previous section and $U=4$.
We use a bath size $L_b=12$ here, for which
the corresponding results for the double occupation have already 
been shown in \figref{figNEQDMFT_U4}(a). 
The symmetry $i\Lambda^{>}(t,t') = i\Lambda^{>}(t+s,t'+s)$ 
is obvious already from the color plot in \figref{figRelaxLambda}(a). 
\figref{figRelaxLambda}(b) then studies a computation based
on the extrapolated $\Lambda\tl{relax}(t,t')$ for which the self-consistency
has only been computed for times $\min(t,t')<t\tl{relax}=1/v_0$. As the difference
to the exact computation is not perceivable with the eye, we show 
a color plot of the difference $i\Lambda\tl{relax}^{>}(t,t') - i\Lambda^{>}(t,t')$.
In particular for times close to the diagonal $t\sim t'$, this difference is almost zero,
but also for the off-diagonal elements, it remains small.

\begin{figure}
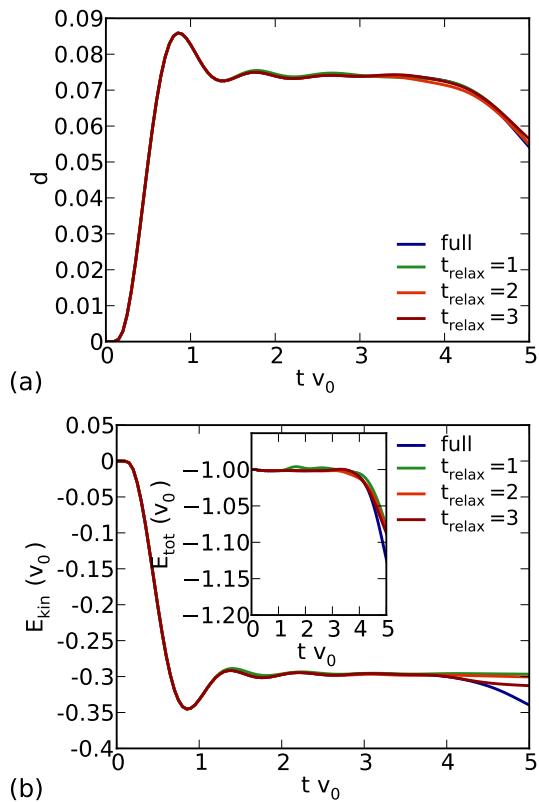

\ig{\numA}{pdf}{\figdirOld fig_offset_d}
\ig{\numA}{pdf}{\figdirOld fig_offset_Ekin_Etot}
\caption{(Color online) 
Comparison between calculation based on the approximated $\Lambda(t,t')$,
for which self-consistency has only been computed for $\min(t,t') < t_\tx{relax}$,
and the exact calculation.}
\label{figRelax}
\end{figure}

\figref{figRelax} shows results for the double occupation and kinetic and total energies
that have been computed using the above described approximation, 
and considers different values for the relaxation time $t\tl{relax}$. Already for the smallest
value studied, $t\tl{relax}=1/v_0$, the result for the double occupation is very close
to the computation that solved the full DMFT loop. For higher values of $t\tl{relax}$, the
approximation converges to the result for which the self-consistency has been solved
in the full $t$-$t'$-plane.

It should be interesting to study theoretically how $t\tl{relax}$
depends on the initial state, quench setup and system parameters, as it measures
the time the system takes until it restores the equilibrium property of the 
most fundamental system parametrization, 
namely the underlying $\Lambda(t,t')$, 
although other observables like the double occupation $d(t)$ might not yet have
relaxed (compare \figsref{figRelax} and \ref{figRelaxLambda}). 
Aside from this theoretical interest, the described extrapolation scheme
helps to speedup computations significantly.

The line of argumentation above should not only apply to quenches to constant Hamiltonians. 
Also for periodically driven system controlled extrapolations should be possible.
In this case a constant extrapolation is no longer appropriate, but 
\eg \tit{linear prediction} can extrapolate regular oscillations with high precision.\cp{numrec07}

\section{Conclusion}
\label{secCon}

For equilibrium DMFT calculations, up to now, it has been difficult for DMRG/MPS-based methods 
to reach the computational efficiencies of CTQMC and NRG computations.\cp{nishimoto04,nishimoto04i,karski05,karski08,ganahl14,wolf14,ganahl14i} 
This situation should be drastically be improved
in view of the results of \secref{secEQ} of this paper, 
where we showed a tremendous speedup of more than 
two orders of magnitude (\figref{figSIAM_chainStar_neq2}) upon using the star 
instead of the chain geometry for the DMFT impurity problem. 
In addition to this speedup, 
the star geometry has tremendous technical advantages when studying 
impurity problems with more than two bands. 
While the chain representation then runs into so-called \tit{normalization} problems 
for the underlyling MPS structure, this is not the case for star geometry.
It now seems feasable to attack the first three-band model within DMFT using an MPS 
based description.

In nonequilibrium, the situation is very different as there is no such 
disadvantage in computational efficiency for DMRG/MPS-based calculations. 
This is due to the fact that parallelization is easily feasable 
for the DMRG computations and one is not so much interested in the behavior of Green functions 
at $t\rightarrow\infty$ (compare \secref{secIntroEQ}). CTQMC, by contrast, then 
has the \tit{phase} problem as a big disadvantage. 
Although NRG has a time-dependent formulation,\cp{anders05}
and despite further progress,\cp{nghiem14} it  
has not yet been employed to treat non-equilibrium DMFT, 
as is the case for the recent numerical operator method.\cp{wang14}

In this paper, we showed that the performance of MPS based computations in the star 
geometry largely exceeds that of exact diagonalization, and bath sizes could be reached
that now make it possible to study more complicated 
setups than quenches from the non-correlated atomic limit.

\section{Acknowledgements}

FAW and US acknowledge
fruitful discussions with M. Eckstein
and support by the research unit FOR 1807 of the 
Deutsche Forschungsgemeinschaft (DFG).

\appendix

\section{Equilbrium DMFT: comparison of star and chain geometries}
\label{secGeoApp}
\subsection{Bath discretization}
\label{secBathDiscretization}

The discrete approximative representation of a SIAM in the star geometry with given bath spectral function
$-\frac{1}{\pi} \tx{Im} \Lambda(\w)$ is given by Hamiltonian \eqref{eqHstar}, where
the parameters in $H\tl{bath}$ and  $H\tl{coupl}$ are\cp{bulla08,wolf14}
\eq{
V_l^2 & = \int_{I_l} d\w\, \big( -\tfrac{1}{\pi} \tx{Im} \Lambda(\w)\big), \quad \\
\e_l  & = \frac{1}{V_l^2} \int_{I_l} d\w\, \w \big( -\tfrac{1}{\pi} \tx{Im} \Lambda(\w) \big). \non
}
Here, the bath discretization intervals are defined as $I_l = [\w_l, \w_{l+1}]$, 
and $\cup_l I_l$ should contain the support of $-\tfrac{1}{\pi} \tx{Im} \Lambda(\w)$. 
We use a linear discretization to define $\{\w_l\}$, 
but a logarithmic discretization can as well be employed.
The creation operators $c_{l\s}\dag$ in \Eqref{eqHstar} can 
be associated with excitations in a certain 
energy interval $I_{l}$ of the bath spectral 
function $-\frac{1}{\pi} \tx{Im} \Lambda(\w)$.

\subsection{Map from star to chain}
\label{secMapStarChain}

Denote the bath orbital (single-particle) states  of the star as $\ket{c_l}$. These are associated 
with the operators $c_{l\s}\dag$ in \Eqref{eqHstar} 
via $\ket{c_l} = c_{l\s}\dag\ket{\tx{vac}}$ (we dropped the spin index in $\ket{c_l}$). 
The first orbital of the chain is then defined as
\eq{ \lb{eqFirstChainState}
\ket{\wt c_{1}} = \frac{1}{\wt V_0} \sum_{l=1}^{L_b} V_{l} \ket{c_{l}} , \qquad \wt V_0 =
\sqrt{\sum_{l}  \abs{V_{l}}^2}.
}
It is a superposition of all states in the star.
$H\tl{hyb}$ in \eqref{eqHhyb} can then be written as  
$H_\tx{hyb} 
= \sum_{\s} \wt V_0 ( \ket{c_{0\s}}\bra{\wt c_{1\s}} + \tx{h.c.})$. 
The Lanczos algorithm constructs a three-diagonal representation 
of $H\tl{bath}+H\tl{hyb}$ by representing it 
in its Gram-Schmidt orthogonalized Krylov basis $\{\ket{\wt c_n}\}$.
$H\tl{hyb}$ is already diagonal in this basis
as by definition it has its single non-zero component for 
$\bra{\wt c_1} H\tl{hyb} \ket{\wt c_1}$, and can be ignored 
for the Lanczos recursion:
\eqs{
\wt \e_n
& = \bra{\wt c_n} H\tl{bath} \ket{\wt c_n}, \\
\ket{r_{n}}
& = H\tl{bath} \ket{\wt c_n} - \wt \e_n \ket{\wt c_n} - \wt V_{n-1} \ket{\wt c_{n-1}}  \\
\wt V_{n}
& = \abs{\ip{r_{n}}{r_{n}}}^{\frac{1}{2}}, \\
\ket{\wt c_{n+1}}
& = \frac{1}{\wt V_{n}} \ket{r_n}, \quad \tx{for}~n = 2, \dots,  L_b-1.
}
For $n=1$, only the definition of $\ket{r_n}$ changes
\eq{
\ket{r_{1}}
  = H\tl{bath} \ket{\wt c_1} - \wt \e_1 \ket{\wt c_1}.
}
The above equations are easily solved by multiplying 
from the left with $\bra{c_l}$ and inserting 
identities $\sum_{l'} \ket{c_{l'}}\bra{c_{l'}}$ such that
the initial vector can be written as 
$(\ip{c_l}{\wt c_{1}})_{l=1}^{L^b} = (V_l)_{l=1}^{L_b}$
and the representation of $H\tl{bath}$ involved is 
$\bra{c_l} H\tl{bath} \ket{c_{l'}} = \e_l \d_{ll'}$. Due to the numerical
instability of the Lanczos algorithm, the recursion has to be 
computed with high-precision arithmetics.

The unitary transform that connects the two geometries
via 
$U\dag ({\cal H}\tl{bath} + {\cal H}\tl{hyb}) U = {\cal H}\tl{pot} + {\cal H}\tl{kin}$,
where 
$({\cal H}\tl{bath} + {\cal H}\tl{hyb})_{ll'} = 
\bra{c_l} H\tl{bath} + H\tl{hyb}\ket{c_{l'}}$ 
and
$({\cal H}\tl{pot} + {\cal H}\tl{kin})_{nn'} = 
\bra{\wt c_n} H\tl{pot} + H\tl{kin}\ket{\wt c_{n'}}$ is given by 
\eq{
(U)_{l,n=1}^{L_b} = (\ip{c_l}{\wt c_n})_{l,n=1}^{L_b} = 
\left(
\begin{array}{ccc}
V_1/\wt V_0      & \ip{c_1}{\wt c_2} & \cdots \\
V_2/\wt V_0      & \ip{c_2}{\wt c_2} & \cdots\\
\vdots   & \vdots  &\\ 
V_{L_b}/\wt V_0  &    &\\
\end{array}
\right)
}
and relates the two basis sets via  
$\ket{\wt c_n} = \sum_l U_{nl}\dag \ket{c_l}$.

\subsection{Time evolution for different interaction strengths}
\label{secGeosInt}

\begin{figure}
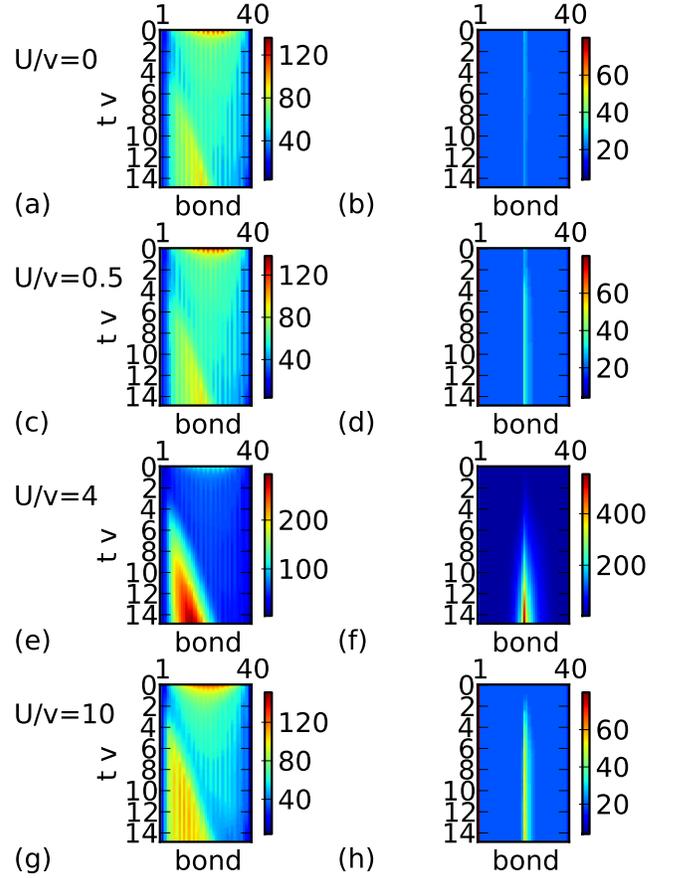

\ig{0.48}{pdf}{\figdirNew fig_SIAM_evolU0_m}
\ig{0.48}{pdf}{\figdirNew fig_SIAM_evolU0.5_m}
\ig{0.48}{pdf}{\figdirNew fig_SIAM_evolU4_m}
\ig{0.48}{pdf}{\figdirNew fig_SIAM_evolU10_m}
\caption{(Color online) 
Bond dimension $m$ versus time and bond
for different interactions strengths $U$ as given in the figure.
We compare the chain geometry (i) (panels (a), (c), (e), (g))
and the star geometry with the impurity located at the center (ii) (panels (b), (d), (f), (h)).
The geometries are defined in \figref{figGeosSketch}.
}
\label{figGeosIntm}
\end{figure}

In \figref{figGeosIntm}, we study the time evolution 
of a SIAM with semi-elliptic bath spectral function \eqref{eqSemiEll} 
for different interaction strengths $U/v \in \{0,0.5,4,10\}$ in the chain geometry (i)
and the star geometry (ii), where the detailed setups are sketched in \figref{figGeosSketch}.
As discussed in \secref{secEntGrowth}, we observe a strongly reduced entanglement
growth in the case of the star geometry in both limits of weak and strong interaction.

\section{Results for a non-self-consistent impurity problem}
\label{secSimpleImp}
\begin{figure}
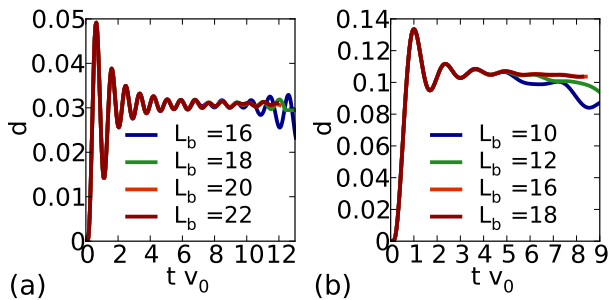

\ig{\numB}{pdf}{\figdirOld fig_BalzerU10}
\ig{\numB}{pdf}{\figdirOld fig_BalzerU4}
\caption{(Color online) 
Results for the solution of the impurity model 
with the hybridization function defined in \Eqref{eqLambdaSimple}
for two interactions $U=10$ (panel (a)) and $U=4$ (panel (b)).
This does not involve the solution of a full self-consistent computation.}
\label{figBalzer}
\end{figure}

Here, we compute the time evolution of a SIAM 
with a hybridization function $\Lambda$ given as
\begin{align}
	\label{eqLambdaSimple}
	\Lambda(t,t')
=v(t)g(t,t')v(t'),\quad t,t'\le t_\text{max},
\end{align}
where $g(t,t') = \theta(t,t') g^>(t,t') + \theta (t',t) g^<(t,t')$ with 
\begin{align}
\label{eq:bathgf}
g^\gtrless(t,t')=\mp i \int d\omega f^\gtrless(\omega)A(\omega)\mathrm{e}^{-\mathrm{i}\omega(t-t')}.
\end{align}
Here, $f^<(\omega)=f(\omega)=1/(\mathrm{e}^{\beta\omega}+1)$, $f^>(\omega)=1-f(\omega)$, and 
the semi-elliptic density of states $A(\omega)=\tfrac{1}{2\pi}\sqrt{4-\omega^2}$. 
We use the temperature $T=1=1/\beta$ and the quench 
from the atomic limit defined in \Eqref{eqRamp}. 
This is the same setup as studied by \ct{balzer14}. 

\figsref{figBalzer}(a) and (b) show results for the double occupation $d(t)$ obtained for 
two different interaction strengths $U/v_0=10$ 
and $U/v_0=4$. 
For the two biggest bath sizes studied, $L_b=20$ and $L_b=22$,
the value for the double occupation agrees up to times $t\sim11/v_0$. 
This has been used by \ct{balzer14} as indicator for that the computation is
controlled. But while \ct{balzer14} could only treat bath sizes up to $L_b=16$, we 
are able to perform controlled computations with bath sizes up to $L_b=24$,
as has already been shown in \figref{figNEQDMFT_U10}, although,
for the full self-consistent calculation, accessible times are much lower 
than here in \figref{figBalzer}. Also, we are able to efficiently treat 
the case $U=4$, which is much more entangled. This limits the efficiency
of multiconfiguration time-dependent Hartree as well as any MPS representation,
but seems to be more severe in the former case.

\section{Regaugeing potentials and couplings does not influence entanglement}
\label{secGauge}
\begin{figure}
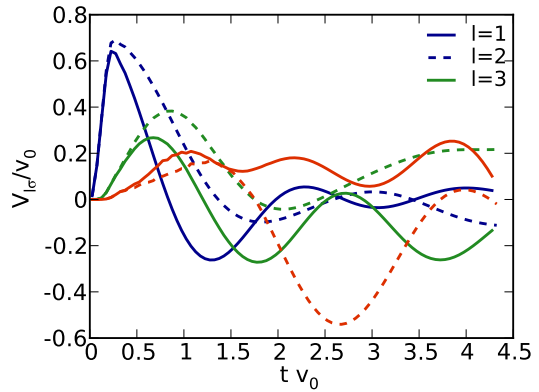

\ig{\numA}{pdf}{\figdirOld fig_regauge_coupl}
\caption{(Color online) 
Time evolution of couplings for the case with homogeneous potential $\e_{l\s}=0$ (full lines) 
and the case with $\e_{l\s} = -2v_0 + l 4v_0/L_b$, where $l$ runs over $L_b$ bath sites.
}
\label{figRegauge}
\end{figure}

In principle, one is free to choose the potentials of the star geometry
arbitrarily if at the same time, one rescales the couplings:\cp{gramsch13}  
Instead of the homogeneous star with time-independent potentials $\e_p$,
which we considered in the previous sections as was done by \ct{gramsch13} and \ct{balzer14},
one can equivalently solve an inhomogeneous star 
with time-independent potentials $\e_p'$.
Instead of reformulating the matrix decomposition with 
a then oscillating non-interacting
Green function $g(t,t',\e_p')\propto e^{-i \e_p' (t-t')}$, one can 
obtain the couplings $V_{p}'(t)$ that correspond to the inhomogeneous model
by a simple gauge transformation of the couplings that correpsond to $\e_p=0$:  
$V_{p}'(t) = V_{p}(t)\exp(-i \epsilon_p' t)$. 

We investigated the question of whether this freedom can be used to 
influence the entanglement properties of the system, starting from the assumption
that the motion of particles in the inhomogeneous system is more constrained 
than in the homogeneous system. This fact is however compensated by the fact that
the couplings $V_p'(t)$ then oscillate much slowlier (see \figref{figRegauge}), which leads to a stronger driving
of the system. 
Equivalent entanglement properties result. This could have been expected,
 as the gauge transformation is a \tit{local} transformation.


\begin{thebibliography}{44}%
\makeatletter
\providecommand \@ifxundefined [1]{%
 \@ifx{#1\undefined}
}%
\providecommand \@ifnum [1]{%
 \ifnum #1\expandafter \@firstoftwo
 \else \expandafter \@secondoftwo
 \fi
}%
\providecommand \@ifx [1]{%
 \ifx #1\expandafter \@firstoftwo
 \else \expandafter \@secondoftwo
 \fi
}%
\providecommand \natexlab [1]{#1}%
\providecommand \enquote  [1]{``#1''}%
\providecommand \bibnamefont  [1]{#1}%
\providecommand \bibfnamefont [1]{#1}%
\providecommand \citenamefont [1]{#1}%
\providecommand \href@noop [0]{\@secondoftwo}%
\providecommand \href [0]{\begingroup \@sanitize@url \@href}%
\providecommand \@href[1]{\@@startlink{#1}\@@href}%
\providecommand \@@href[1]{\endgroup#1\@@endlink}%
\providecommand \@sanitize@url [0]{\catcode `\\12\catcode `\$12\catcode
  `\&12\catcode `\#12\catcode `\^12\catcode `\_12\catcode `\%12\relax}%
\providecommand \@@startlink[1]{}%
\providecommand \@@endlink[0]{}%
\providecommand \url  [0]{\begingroup\@sanitize@url \@url }%
\providecommand \@url [1]{\endgroup\@href {#1}{\urlprefix }}%
\providecommand \urlprefix  [0]{URL }%
\providecommand \Eprint [0]{\href }%
\providecommand \doibase [0]{http://dx.doi.org/}%
\providecommand \selectlanguage [0]{\@gobble}%
\providecommand \bibinfo  [0]{\@secondoftwo}%
\providecommand \bibfield  [0]{\@secondoftwo}%
\providecommand \translation [1]{[#1]}%
\providecommand \BibitemOpen [0]{}%
\providecommand \bibitemStop [0]{}%
\providecommand \bibitemNoStop [0]{.\EOS\space}%
\providecommand \EOS [0]{\spacefactor3000\relax}%
\providecommand \BibitemShut  [1]{\csname bibitem#1\endcsname}%
\let\auto@bib@innerbib\@empty
\bibitem [{\citenamefont {Metzner}\ and\ \citenamefont
  {Vollhardt}(1989)}]{metzner89}%
  \BibitemOpen
  \bibfield  {author} {\bibinfo {author} {\bibfnamefont {W.}~\bibnamefont
  {Metzner}}\ and\ \bibinfo {author} {\bibfnamefont {D.}~\bibnamefont
  {Vollhardt}},\ }\href {\doibase 10.1103/PhysRevLett.62.324} {\bibfield
  {journal} {\bibinfo  {journal} {Physical Review Letters}\ }\textbf {\bibinfo
  {volume} {62}},\ \bibinfo {pages} {324} (\bibinfo {year} {1989})}\BibitemShut
  {NoStop}%
\bibitem [{\citenamefont {Georges}\ and\ \citenamefont
  {Kotliar}(1992)}]{georges92}%
  \BibitemOpen
  \bibfield  {author} {\bibinfo {author} {\bibfnamefont {A.}~\bibnamefont
  {Georges}}\ and\ \bibinfo {author} {\bibfnamefont {G.}~\bibnamefont
  {Kotliar}},\ }\href {\doibase 10.1103/PhysRevB.45.6479} {\bibfield  {journal}
  {\bibinfo  {journal} {Phys. Rev. B}\ }\textbf {\bibinfo {volume} {45}},\
  \bibinfo {pages} {6479} (\bibinfo {year} {1992})}\BibitemShut {NoStop}%
\bibitem [{\citenamefont {Georges}\ \emph {et~al.}(1996)\citenamefont
  {Georges}, \citenamefont {Kotliar}, \citenamefont {Krauth},\ and\
  \citenamefont {Rozenberg}}]{georges96}%
  \BibitemOpen
  \bibfield  {author} {\bibinfo {author} {\bibfnamefont {A.}~\bibnamefont
  {Georges}}, \bibinfo {author} {\bibfnamefont {G.}~\bibnamefont {Kotliar}},
  \bibinfo {author} {\bibfnamefont {W.}~\bibnamefont {Krauth}}, \ and\ \bibinfo
  {author} {\bibfnamefont {M.~J.}\ \bibnamefont {Rozenberg}},\ }\href {\doibase
  10.1103/RevModPhys.68.13} {\bibfield  {journal} {\bibinfo  {journal} {Rev.
  Mod. Phys.}\ }\textbf {\bibinfo {volume} {68}},\ \bibinfo {pages} {13}
  (\bibinfo {year} {1996})}\BibitemShut {NoStop}%
\bibitem [{\citenamefont {Kotliar}\ \emph {et~al.}(2006)\citenamefont
  {Kotliar}, \citenamefont {Savrasov}, \citenamefont {Haule}, \citenamefont
  {Oudovenko}, \citenamefont {Parcollet},\ and\ \citenamefont
  {Marianetti}}]{kotliar06}%
  \BibitemOpen
  \bibfield  {author} {\bibinfo {author} {\bibfnamefont {G.}~\bibnamefont
  {Kotliar}}, \bibinfo {author} {\bibfnamefont {S.}~\bibnamefont {Savrasov}},
  \bibinfo {author} {\bibfnamefont {K.}~\bibnamefont {Haule}}, \bibinfo
  {author} {\bibfnamefont {V.}~\bibnamefont {Oudovenko}}, \bibinfo {author}
  {\bibfnamefont {O.}~\bibnamefont {Parcollet}}, \ and\ \bibinfo {author}
  {\bibfnamefont {C.}~\bibnamefont {Marianetti}},\ }\href {\doibase
  10.1103/RevModPhys.78.865} {\bibfield  {journal} {\bibinfo  {journal}
  {Reviews of Modern Physics}\ }\textbf {\bibinfo {volume} {78}},\ \bibinfo
  {pages} {865} (\bibinfo {year} {2006})}\BibitemShut {NoStop}%
\bibitem [{\citenamefont {Schmidt}\ and\ \citenamefont
  {Monien}(2002)}]{schmidt02}%
  \BibitemOpen
  \bibfield  {author} {\bibinfo {author} {\bibfnamefont {P.}~\bibnamefont
  {Schmidt}}\ and\ \bibinfo {author} {\bibfnamefont {H.}~\bibnamefont
  {Monien}},\ }\href {http://arxiv.org/abs/cond-mat/0202046} {\bibfield
  {journal} {\bibinfo  {journal} {ArXiv}\ ,\ \bibinfo {pages} {cond}} (\bibinfo
  {year} {2002})},\ \Eprint {http://arxiv.org/abs/cond-mat/0202046}
  {cond-mat/0202046} \BibitemShut {NoStop}%
\bibitem [{\citenamefont {Freericks}\ \emph {et~al.}(2006)\citenamefont
  {Freericks}, \citenamefont {Turkowski},\ and\ \citenamefont
  {Zlati{\'c}}}]{freericks06}%
  \BibitemOpen
  \bibfield  {author} {\bibinfo {author} {\bibfnamefont {J.~K.}\ \bibnamefont
  {Freericks}}, \bibinfo {author} {\bibfnamefont {V.~M.}\ \bibnamefont
  {Turkowski}}, \ and\ \bibinfo {author} {\bibfnamefont {V.}~\bibnamefont
  {Zlati{\'c}}},\ }\href {\doibase 10.1103/physrevlett.97.266408} {\bibfield
  {journal} {\bibinfo  {journal} {Physical Review Letters}\ }\textbf {\bibinfo
  {volume} {97}},\ \bibinfo {pages} {266408} (\bibinfo {year}
  {2006})}\BibitemShut {NoStop}%
\bibitem [{\citenamefont {Aoki}\ \emph {et~al.}(2014)\citenamefont {Aoki},
  \citenamefont {Tsuji}, \citenamefont {Eckstein}, \citenamefont {Kollar},
  \citenamefont {Oka},\ and\ \citenamefont {Werner}}]{aoki14}%
  \BibitemOpen
  \bibfield  {author} {\bibinfo {author} {\bibfnamefont {H.}~\bibnamefont
  {Aoki}}, \bibinfo {author} {\bibfnamefont {N.}~\bibnamefont {Tsuji}},
  \bibinfo {author} {\bibfnamefont {M.}~\bibnamefont {Eckstein}}, \bibinfo
  {author} {\bibfnamefont {M.}~\bibnamefont {Kollar}}, \bibinfo {author}
  {\bibfnamefont {T.}~\bibnamefont {Oka}}, \ and\ \bibinfo {author}
  {\bibfnamefont {P.}~\bibnamefont {Werner}},\ }\href {\doibase
  10.1103/revmodphys.86.779} {\bibfield  {journal} {\bibinfo  {journal} {Rev.
  Mod. Phys.}\ }\textbf {\bibinfo {volume} {86}},\ \bibinfo {pages} {779}
  (\bibinfo {year} {2014})}\BibitemShut {NoStop}%
\bibitem [{\citenamefont {Eckstein}\ \emph
  {et~al.}(2009{\natexlab{a}})\citenamefont {Eckstein}, \citenamefont
  {Kollar},\ and\ \citenamefont {Werner}}]{eckstein09}%
  \BibitemOpen
  \bibfield  {author} {\bibinfo {author} {\bibfnamefont {M.}~\bibnamefont
  {Eckstein}}, \bibinfo {author} {\bibfnamefont {M.}~\bibnamefont {Kollar}}, \
  and\ \bibinfo {author} {\bibfnamefont {P.}~\bibnamefont {Werner}},\ }\href
  {\doibase 10.1103/PhysRevLett.103.056403} {\bibfield  {journal} {\bibinfo
  {journal} {Phys. Rev. Lett.}\ }\textbf {\bibinfo {volume} {103}},\ \bibinfo
  {pages} {056403} (\bibinfo {year} {2009}{\natexlab{a}})}\BibitemShut
  {NoStop}%
\bibitem [{\citenamefont {Eckstein}\ and\ \citenamefont
  {Werner}(2010)}]{eckstein10}%
  \BibitemOpen
  \bibfield  {author} {\bibinfo {author} {\bibfnamefont {M.}~\bibnamefont
  {Eckstein}}\ and\ \bibinfo {author} {\bibfnamefont {P.}~\bibnamefont
  {Werner}},\ }\href {\doibase 10.1103/physrevb.82.115115} {\bibfield
  {journal} {\bibinfo  {journal} {Phys. Rev. B}\ }\textbf {\bibinfo {volume}
  {82}},\ \bibinfo {pages} {115115} (\bibinfo {year} {2010})}\BibitemShut
  {NoStop}%
\bibitem [{\citenamefont {Eckstein}\ \emph {et~al.}(2010)\citenamefont
  {Eckstein}, \citenamefont {Kollar},\ and\ \citenamefont
  {Werner}}]{eckstein10i}%
  \BibitemOpen
  \bibfield  {author} {\bibinfo {author} {\bibfnamefont {M.}~\bibnamefont
  {Eckstein}}, \bibinfo {author} {\bibfnamefont {M.}~\bibnamefont {Kollar}}, \
  and\ \bibinfo {author} {\bibfnamefont {P.}~\bibnamefont {Werner}},\ }\href
  {\doibase 10.1103/physrevb.81.115131} {\bibfield  {journal} {\bibinfo
  {journal} {Phys. Rev. B}\ }\textbf {\bibinfo {volume} {81}},\ \bibinfo
  {pages} {115131} (\bibinfo {year} {2010})}\BibitemShut {NoStop}%
\bibitem [{\citenamefont {Amaricci}\ \emph {et~al.}(2012)\citenamefont
  {Amaricci}, \citenamefont {Weber}, \citenamefont {Capone},\ and\
  \citenamefont {Kotliar}}]{amaricci12}%
  \BibitemOpen
  \bibfield  {author} {\bibinfo {author} {\bibfnamefont {A.}~\bibnamefont
  {Amaricci}}, \bibinfo {author} {\bibfnamefont {C.}~\bibnamefont {Weber}},
  \bibinfo {author} {\bibfnamefont {M.}~\bibnamefont {Capone}}, \ and\ \bibinfo
  {author} {\bibfnamefont {G.}~\bibnamefont {Kotliar}},\ }\href {\doibase
  10.1103/physrevb.86.085110} {\bibfield  {journal} {\bibinfo  {journal} {Phys.
  Rev. B}\ }\textbf {\bibinfo {volume} {86}},\ \bibinfo {pages} {085110}
  (\bibinfo {year} {2012})}\BibitemShut {NoStop}%
\bibitem [{\citenamefont {Tsuji}\ and\ \citenamefont {Werner}(2013)}]{tsuji13}%
  \BibitemOpen
  \bibfield  {author} {\bibinfo {author} {\bibfnamefont {N.}~\bibnamefont
  {Tsuji}}\ and\ \bibinfo {author} {\bibfnamefont {P.}~\bibnamefont {Werner}},\
  }\href {\doibase 10.1103/physrevb.88.165115} {\bibfield  {journal} {\bibinfo
  {journal} {Phys. Rev. B}\ }\textbf {\bibinfo {volume} {88}},\ \bibinfo
  {pages} {165115} (\bibinfo {year} {2013})}\BibitemShut {NoStop}%
\bibitem [{\citenamefont {Arrigoni}\ \emph {et~al.}(2013)\citenamefont
  {Arrigoni}, \citenamefont {Knap},\ and\ \citenamefont {von~der
  Linden}}]{arrigoni13}%
  \BibitemOpen
  \bibfield  {author} {\bibinfo {author} {\bibfnamefont {E.}~\bibnamefont
  {Arrigoni}}, \bibinfo {author} {\bibfnamefont {M.}~\bibnamefont {Knap}}, \
  and\ \bibinfo {author} {\bibfnamefont {W.}~\bibnamefont {von~der Linden}},\
  }\href {\doibase 10.1103/PhysRevLett.110.086403} {\bibfield  {journal}
  {\bibinfo  {journal} {Physical Review Letters}\ }\textbf {\bibinfo {volume}
  {110}},\ \bibinfo {pages} {086403} (\bibinfo {year} {2013})}\BibitemShut
  {NoStop}%
\bibitem [{\citenamefont {Gramsch}\ \emph {et~al.}(2013)\citenamefont
  {Gramsch}, \citenamefont {Balzer}, \citenamefont {Eckstein},\ and\
  \citenamefont {Kollar}}]{gramsch13}%
  \BibitemOpen
  \bibfield  {author} {\bibinfo {author} {\bibfnamefont {C.}~\bibnamefont
  {Gramsch}}, \bibinfo {author} {\bibfnamefont {K.}~\bibnamefont {Balzer}},
  \bibinfo {author} {\bibfnamefont {M.}~\bibnamefont {Eckstein}}, \ and\
  \bibinfo {author} {\bibfnamefont {M.}~\bibnamefont {Kollar}},\ }\href
  {\doibase 10.1103/physrevb.88.235106} {\bibfield  {journal} {\bibinfo
  {journal} {Phys. Rev. B}\ }\textbf {\bibinfo {volume} {88}},\ \bibinfo
  {pages} {235106} (\bibinfo {year} {2013})}\BibitemShut {NoStop}%
\bibitem [{\citenamefont {Balzer}\ and\ \citenamefont
  {Eckstein}(2014)}]{balzer14i}%
  \BibitemOpen
  \bibfield  {author} {\bibinfo {author} {\bibfnamefont {K.}~\bibnamefont
  {Balzer}}\ and\ \bibinfo {author} {\bibfnamefont {M.}~\bibnamefont
  {Eckstein}},\ }\href {\doibase 10.1103/physrevb.89.035148} {\bibfield
  {journal} {\bibinfo  {journal} {Phys. Rev. B}\ }\textbf {\bibinfo {volume}
  {89}},\ \bibinfo {pages} {035148} (\bibinfo {year} {2014})}\BibitemShut
  {NoStop}%
\bibitem [{\citenamefont {Schollw\"ock}(2005)}]{schollwock05}%
  \BibitemOpen
  \bibfield  {author} {\bibinfo {author} {\bibfnamefont {U.}~\bibnamefont
  {Schollw\"ock}},\ }\href {\doibase 10.1103/RevModPhys.77.259} {\bibfield
  {journal} {\bibinfo  {journal} {Rev. Mod. Phys.}\ }\textbf {\bibinfo {volume}
  {77}},\ \bibinfo {pages} {259} (\bibinfo {year} {2005})}\BibitemShut
  {NoStop}%
\bibitem [{\citenamefont {Schollw\"ock}(2011)}]{schollwock11}%
  \BibitemOpen
  \bibfield  {author} {\bibinfo {author} {\bibfnamefont {U.}~\bibnamefont
  {Schollw\"ock}},\ }\href {\doibase 10.1016/j.aop.2010.09.012} {\bibfield
  {journal} {\bibinfo  {journal} {Annals of Physics}\ }\textbf {\bibinfo
  {volume} {326}},\ \bibinfo {pages} {96} (\bibinfo {year} {2011})}\BibitemShut
  {NoStop}%
\bibitem [{\citenamefont {Verstraete}\ and\ \citenamefont
  {Cirac}(2004)}]{verstraete04}%
  \BibitemOpen
  \bibfield  {author} {\bibinfo {author} {\bibfnamefont {F.}~\bibnamefont
  {Verstraete}}\ and\ \bibinfo {author} {\bibfnamefont {J.~I.}\ \bibnamefont
  {Cirac}},\ }\href {http://arxiv.org/abs/cond-mat/0407066} {\bibfield
  {journal} {\bibinfo  {journal} {ArXiv}\ ,\ \bibinfo {pages} {cond}} (\bibinfo
  {year} {2004})},\ \Eprint {http://arxiv.org/abs/cond-mat/0407066}
  {cond-mat/0407066} \BibitemShut {NoStop}%
\bibitem [{\citenamefont {Vidal}(2007)}]{vidal07}%
  \BibitemOpen
  \bibfield  {author} {\bibinfo {author} {\bibfnamefont {G.}~\bibnamefont
  {Vidal}},\ }\href {\doibase 10.1103/physrevlett.99.220405} {\bibfield
  {journal} {\bibinfo  {journal} {Physical Review Letters}\ }\textbf {\bibinfo
  {volume} {99}},\ \bibinfo {pages} {220405} (\bibinfo {year}
  {2007})}\BibitemShut {NoStop}%
\bibitem [{\citenamefont {Balzer}\ \emph {et~al.}(2014)\citenamefont {Balzer},
  \citenamefont {Li}, \citenamefont {Vendrell},\ and\ \citenamefont
  {Eckstein}}]{balzer14}%
  \BibitemOpen
  \bibfield  {author} {\bibinfo {author} {\bibfnamefont {K.}~\bibnamefont
  {Balzer}}, \bibinfo {author} {\bibfnamefont {Z.}~\bibnamefont {Li}}, \bibinfo
  {author} {\bibfnamefont {O.}~\bibnamefont {Vendrell}}, \ and\ \bibinfo
  {author} {\bibfnamefont {M.}~\bibnamefont {Eckstein}},\ }\href
  {http://arxiv.org/abs/1407.6578} {\bibfield  {journal} {\bibinfo  {journal}
  {ArXiv}\ ,\ \bibinfo {pages} {1407.6578}} (\bibinfo {year}
  {2014})}\BibitemShut {NoStop}%
\bibitem [{\citenamefont {Keldysh}(1964)}]{keldysh64}%
  \BibitemOpen
  \bibfield  {author} {\bibinfo {author} {\bibfnamefont {L.~V.}\ \bibnamefont
  {Keldysh}},\ }\href@noop {} {\bibfield  {journal} {\bibinfo  {journal} {Zh.
  Eksp. Teor. Fiz.}\ }\textbf {\bibinfo {volume} {47}},\ \bibinfo {pages}
  {1515} (\bibinfo {year} {1964})},\ \bibinfo {note} {[Sov. Phys. JETP 20,1018
  (1965)]}\BibitemShut {NoStop}%
\bibitem [{\citenamefont {Eckstein}\ \emph
  {et~al.}(2009{\natexlab{b}})\citenamefont {Eckstein}, \citenamefont {Hackl},
  \citenamefont {Kehrein}, \citenamefont {Kollar}, \citenamefont {Moeckel},
  \citenamefont {Werner},\ and\ \citenamefont {Wolf}}]{eckstein09i}%
  \BibitemOpen
  \bibfield  {author} {\bibinfo {author} {\bibfnamefont {M.}~\bibnamefont
  {Eckstein}}, \bibinfo {author} {\bibfnamefont {A.}~\bibnamefont {Hackl}},
  \bibinfo {author} {\bibfnamefont {S.}~\bibnamefont {Kehrein}}, \bibinfo
  {author} {\bibfnamefont {M.}~\bibnamefont {Kollar}}, \bibinfo {author}
  {\bibfnamefont {M.}~\bibnamefont {Moeckel}}, \bibinfo {author} {\bibfnamefont
  {P.}~\bibnamefont {Werner}}, \ and\ \bibinfo {author} {\bibfnamefont {F.~A.}\
  \bibnamefont {Wolf}},\ }\href {\doibase 10.1140/epjst/e2010-01219-x}
  {\bibfield  {journal} {\bibinfo  {journal} {The European Physical Journal
  Special Topics}\ }\textbf {\bibinfo {volume} {180}},\ \bibinfo {pages} {217}
  (\bibinfo {year} {2009}{\natexlab{b}})}\BibitemShut {NoStop}%
\bibitem [{\citenamefont {Eckstein}(2009)}]{eckstein09thesis}%
  \BibitemOpen
  \bibfield  {author} {\bibinfo {author} {\bibfnamefont {M.}~\bibnamefont
  {Eckstein}},\ }\emph {\bibinfo {title} {Nonequilibrium DMFT}},\ \href@noop {}
  {Ph.D. thesis},\ \bibinfo  {school} {Uni Augsburg} (\bibinfo {year}
  {2009})\BibitemShut {NoStop}%
\bibitem [{\citenamefont {Garc\'ia}\ \emph {et~al.}(2004)\citenamefont
  {Garc\'ia}, \citenamefont {Hallberg},\ and\ \citenamefont
  {Rozenberg}}]{garcia04}%
  \BibitemOpen
  \bibfield  {author} {\bibinfo {author} {\bibfnamefont {D.~J.}\ \bibnamefont
  {Garc\'ia}}, \bibinfo {author} {\bibfnamefont {K.}~\bibnamefont {Hallberg}},
  \ and\ \bibinfo {author} {\bibfnamefont {M.~J.}\ \bibnamefont {Rozenberg}},\
  }\href {\doibase 10.1103/PhysRevLett.93.246403} {\bibfield  {journal}
  {\bibinfo  {journal} {Phys. Rev. Lett.}\ }\textbf {\bibinfo {volume} {93}},\
  \bibinfo {pages} {246403} (\bibinfo {year} {2004})}\BibitemShut {NoStop}%
\bibitem [{\citenamefont {Nishimoto}\ and\ \citenamefont
  {Jeckelmann}(2004)}]{nishimoto04}%
  \BibitemOpen
  \bibfield  {author} {\bibinfo {author} {\bibfnamefont {S.}~\bibnamefont
  {Nishimoto}}\ and\ \bibinfo {author} {\bibfnamefont {E.}~\bibnamefont
  {Jeckelmann}},\ }\href {\doibase 10.1088/0953-8984/16/4/010} {\bibfield
  {journal} {\bibinfo  {journal} {J. Phys.: Condens. Matter}\ }\textbf
  {\bibinfo {volume} {16}},\ \bibinfo {pages} {613} (\bibinfo {year}
  {2004})}\BibitemShut {NoStop}%
\bibitem [{\citenamefont {Nishimoto}\ \emph {et~al.}(2004)\citenamefont
  {Nishimoto}, \citenamefont {Gebhard},\ and\ \citenamefont
  {Jeckelmann}}]{nishimoto04i}%
  \BibitemOpen
  \bibfield  {author} {\bibinfo {author} {\bibfnamefont {S.}~\bibnamefont
  {Nishimoto}}, \bibinfo {author} {\bibfnamefont {F.}~\bibnamefont {Gebhard}},
  \ and\ \bibinfo {author} {\bibfnamefont {E.}~\bibnamefont {Jeckelmann}},\
  }\href {\doibase 10.1088/0953-8984/16/39/038} {\bibfield  {journal} {\bibinfo
   {journal} {J. Phys.: Condens. Matter}\ }\textbf {\bibinfo {volume} {16}},\
  \bibinfo {pages} {7063} (\bibinfo {year} {2004})}\BibitemShut {NoStop}%
\bibitem [{\citenamefont {Karski}\ \emph {et~al.}(2005)\citenamefont {Karski},
  \citenamefont {Raas},\ and\ \citenamefont {Uhrig}}]{karski05}%
  \BibitemOpen
  \bibfield  {author} {\bibinfo {author} {\bibfnamefont {M.}~\bibnamefont
  {Karski}}, \bibinfo {author} {\bibfnamefont {C.}~\bibnamefont {Raas}}, \ and\
  \bibinfo {author} {\bibfnamefont {G.~S.}\ \bibnamefont {Uhrig}},\ }\href
  {\doibase 10.1103/PhysRevB.72.113110} {\bibfield  {journal} {\bibinfo
  {journal} {Phys. Rev. B}\ }\textbf {\bibinfo {volume} {72}},\ \bibinfo
  {pages} {113110} (\bibinfo {year} {2005})}\BibitemShut {NoStop}%
\bibitem [{\citenamefont {Garc\'ia}\ \emph {et~al.}(2007)\citenamefont
  {Garc\'ia}, \citenamefont {Miranda}, \citenamefont {Hallberg},\ and\
  \citenamefont {Rozenberg}}]{garcia07}%
  \BibitemOpen
  \bibfield  {author} {\bibinfo {author} {\bibfnamefont {D.~J.}\ \bibnamefont
  {Garc\'ia}}, \bibinfo {author} {\bibfnamefont {E.}~\bibnamefont {Miranda}},
  \bibinfo {author} {\bibfnamefont {K.}~\bibnamefont {Hallberg}}, \ and\
  \bibinfo {author} {\bibfnamefont {M.~J.}\ \bibnamefont {Rozenberg}},\ }\href
  {\doibase 10.1103/physrevb.75.121102} {\bibfield  {journal} {\bibinfo
  {journal} {Phys. Rev. B}\ }\textbf {\bibinfo {volume} {75}},\ \bibinfo
  {pages} {121102} (\bibinfo {year} {2007})}\BibitemShut {NoStop}%
\bibitem [{\citenamefont {Karski}\ \emph {et~al.}(2008)\citenamefont {Karski},
  \citenamefont {Raas},\ and\ \citenamefont {Uhrig}}]{karski08}%
  \BibitemOpen
  \bibfield  {author} {\bibinfo {author} {\bibfnamefont {M.}~\bibnamefont
  {Karski}}, \bibinfo {author} {\bibfnamefont {C.}~\bibnamefont {Raas}}, \ and\
  \bibinfo {author} {\bibfnamefont {G.~S.}\ \bibnamefont {Uhrig}},\ }\href
  {\doibase 10.1103/PhysRevB.77.075116} {\bibfield  {journal} {\bibinfo
  {journal} {Phys. Rev. B}\ }\textbf {\bibinfo {volume} {77}},\ \bibinfo
  {pages} {075116} (\bibinfo {year} {2008})}\BibitemShut {NoStop}%
\bibitem [{\citenamefont {Ganahl}\ \emph
  {et~al.}(2014{\natexlab{a}})\citenamefont {Ganahl}, \citenamefont
  {Thunstr{\"o}m}, \citenamefont {Verstraete}, \citenamefont {Held},\ and\
  \citenamefont {Evertz}}]{ganahl14}%
  \BibitemOpen
  \bibfield  {author} {\bibinfo {author} {\bibfnamefont {M.}~\bibnamefont
  {Ganahl}}, \bibinfo {author} {\bibfnamefont {P.}~\bibnamefont
  {Thunstr{\"o}m}}, \bibinfo {author} {\bibfnamefont {F.}~\bibnamefont
  {Verstraete}}, \bibinfo {author} {\bibfnamefont {K.}~\bibnamefont {Held}}, \
  and\ \bibinfo {author} {\bibfnamefont {H.~G.}\ \bibnamefont {Evertz}},\
  }\href {\doibase 10.1103/physrevb.90.045144} {\bibfield  {journal} {\bibinfo
  {journal} {Phys. Rev. B}\ }\textbf {\bibinfo {volume} {90}},\ \bibinfo
  {pages} {045144} (\bibinfo {year} {2014}{\natexlab{a}})}\BibitemShut
  {NoStop}%
\bibitem [{\citenamefont {Wolf}\ \emph {et~al.}(2014)\citenamefont {Wolf},
  \citenamefont {McCulloch}, \citenamefont {Parcollet},\ and\ \citenamefont
  {Schollw{\"o}ck}}]{wolf14}%
  \BibitemOpen
  \bibfield  {author} {\bibinfo {author} {\bibfnamefont {F.~A.}\ \bibnamefont
  {Wolf}}, \bibinfo {author} {\bibfnamefont {I.~P.}\ \bibnamefont {McCulloch}},
  \bibinfo {author} {\bibfnamefont {O.}~\bibnamefont {Parcollet}}, \ and\
  \bibinfo {author} {\bibfnamefont {U.}~\bibnamefont {Schollw{\"o}ck}},\ }\href
  {\doibase 10.1103/physrevb.90.115124} {\bibfield  {journal} {\bibinfo
  {journal} {Phys. Rev. B}\ }\textbf {\bibinfo {volume} {90}},\ \bibinfo
  {pages} {115124} (\bibinfo {year} {2014})}\BibitemShut {NoStop}%
\bibitem [{\citenamefont {Ganahl}\ \emph
  {et~al.}(2014{\natexlab{b}})\citenamefont {Ganahl}, \citenamefont {Aichhorn},
  \citenamefont {Thunstr\"om}, \citenamefont {Held}, \citenamefont {Evertz},\
  and\ \citenamefont {Verstraete}}]{ganahl14i}%
  \BibitemOpen
  \bibfield  {author} {\bibinfo {author} {\bibfnamefont {M.}~\bibnamefont
  {Ganahl}}, \bibinfo {author} {\bibfnamefont {M.}~\bibnamefont {Aichhorn}},
  \bibinfo {author} {\bibfnamefont {P.}~\bibnamefont {Thunstr\"om}}, \bibinfo
  {author} {\bibfnamefont {K.}~\bibnamefont {Held}}, \bibinfo {author}
  {\bibfnamefont {H.~G.}\ \bibnamefont {Evertz}}, \ and\ \bibinfo {author}
  {\bibfnamefont {F.}~\bibnamefont {Verstraete}},\ }\href
  {http://arxiv.org/abs/1405.6728} {\bibfield  {journal} {\bibinfo  {journal}
  {preprint}\ } (\bibinfo {year} {2014}{\natexlab{b}})},\ \Eprint
  {http://arxiv.org/abs/1405.6728} {arXiv:1405.6728} \BibitemShut {NoStop}%
\bibitem [{\citenamefont {Raas}\ \emph {et~al.}(2004)\citenamefont {Raas},
  \citenamefont {Uhrig},\ and\ \citenamefont {Anders}}]{raas04}%
  \BibitemOpen
  \bibfield  {author} {\bibinfo {author} {\bibfnamefont {C.}~\bibnamefont
  {Raas}}, \bibinfo {author} {\bibfnamefont {G.~S.}\ \bibnamefont {Uhrig}}, \
  and\ \bibinfo {author} {\bibfnamefont {F.~B.}\ \bibnamefont {Anders}},\
  }\href {\doibase 10.1103/PhysRevB.69.041102} {\bibfield  {journal} {\bibinfo
  {journal} {Phys. Rev. B}\ }\textbf {\bibinfo {volume} {69}},\ \bibinfo
  {pages} {041102} (\bibinfo {year} {2004})}\BibitemShut {NoStop}%
\bibitem [{\citenamefont {Nuss}\ \emph {et~al.}(2014)\citenamefont {Nuss},
  \citenamefont {Ganahl}, \citenamefont {Arrigoni}, \citenamefont {von~der
  Linden},\ and\ \citenamefont {Evertz}}]{nuss14}%
  \BibitemOpen
  \bibfield  {author} {\bibinfo {author} {\bibfnamefont {M.}~\bibnamefont
  {Nuss}}, \bibinfo {author} {\bibfnamefont {M.}~\bibnamefont {Ganahl}},
  \bibinfo {author} {\bibfnamefont {E.}~\bibnamefont {Arrigoni}}, \bibinfo
  {author} {\bibfnamefont {W.}~\bibnamefont {von~der Linden}}, \ and\ \bibinfo
  {author} {\bibfnamefont {H.~G.}\ \bibnamefont {Evertz}},\ }\href@noop {}
  {\bibfield  {journal} {\bibinfo  {journal} {ArXiv}\ ,\ \bibinfo {pages}
  {1409.0646}} (\bibinfo {year} {2014})},\ \Eprint
  {http://arxiv.org/abs/1409.0646} {1409.0646} \BibitemShut {NoStop}%
\bibitem [{\citenamefont {Bulla}\ \emph {et~al.}(2008)\citenamefont {Bulla},
  \citenamefont {Costi},\ and\ \citenamefont {Pruschke}}]{bulla08}%
  \BibitemOpen
  \bibfield  {author} {\bibinfo {author} {\bibfnamefont {R.}~\bibnamefont
  {Bulla}}, \bibinfo {author} {\bibfnamefont {T.}~\bibnamefont {Costi}}, \ and\
  \bibinfo {author} {\bibfnamefont {T.}~\bibnamefont {Pruschke}},\ }\href
  {\doibase 10.1103/RevModPhys.80.395} {\bibfield  {journal} {\bibinfo
  {journal} {Rev. Mod. Phys.}\ }\textbf {\bibinfo {volume} {80}},\ \bibinfo
  {pages} {395} (\bibinfo {year} {2008})}\BibitemShut {NoStop}%
\bibitem [{\citenamefont {Raas}(2005)}]{raas05}%
  \BibitemOpen
  \bibfield  {author} {\bibinfo {author} {\bibfnamefont {C.}~\bibnamefont
  {Raas}},\ }\emph {\bibinfo {title} {Dynamic Density-Matrix Renormalization
  for the Symmetric Single Impurity Anderson Model}},\ \href
  {http://www.raas.de/thesis.html} {Ph.D. thesis},\ \bibinfo  {school}
  {University of Cologne} (\bibinfo {year} {2005})\BibitemShut {NoStop}%
\bibitem [{\citenamefont {Pekker}\ and\ \citenamefont
  {Clark}(2014)}]{pekker14}%
  \BibitemOpen
  \bibfield  {author} {\bibinfo {author} {\bibfnamefont {D.}~\bibnamefont
  {Pekker}}\ and\ \bibinfo {author} {\bibfnamefont {B.~K.}\ \bibnamefont
  {Clark}},\ }\href {http://arxiv.org/abs/1410.2224} {\bibfield  {journal}
  {\bibinfo  {journal} {ArXiv}\ ,\ \bibinfo {pages} {1410.2224}} (\bibinfo
  {year} {2014})},\ \Eprint {http://arxiv.org/abs/1410.2224} {1410.2224}
  \BibitemShut {NoStop}%
\bibitem [{\citenamefont {Lu}\ \emph {et~al.}(2014)\citenamefont {Lu},
  \citenamefont {H{\"o}ppner}, \citenamefont {Gunnarsson},\ and\ \citenamefont
  {Haverkort}}]{lu14}%
  \BibitemOpen
  \bibfield  {author} {\bibinfo {author} {\bibfnamefont {Y.}~\bibnamefont
  {Lu}}, \bibinfo {author} {\bibfnamefont {M.}~\bibnamefont {H{\"o}ppner}},
  \bibinfo {author} {\bibfnamefont {O.}~\bibnamefont {Gunnarsson}}, \ and\
  \bibinfo {author} {\bibfnamefont {M.~W.}\ \bibnamefont {Haverkort}},\ }\href
  {\doibase 10.1103/physrevb.90.085102} {\bibfield  {journal} {\bibinfo
  {journal} {Phys. Rev. B}\ }\textbf {\bibinfo {volume} {90}},\ \bibinfo
  {pages} {085102} (\bibinfo {year} {2014})}\BibitemShut {NoStop}%
\bibitem [{\citenamefont {Press}\ \emph {et~al.}(2007)\citenamefont {Press},
  \citenamefont {Teukolsky}, \citenamefont {Vetterling},\ and\ \citenamefont
  {Flannery}}]{numrec07}%
  \BibitemOpen
  \bibfield  {author} {\bibinfo {author} {\bibfnamefont {W.~H.}\ \bibnamefont
  {Press}}, \bibinfo {author} {\bibfnamefont {S.~A.}\ \bibnamefont
  {Teukolsky}}, \bibinfo {author} {\bibfnamefont {W.~T.}\ \bibnamefont
  {Vetterling}}, \ and\ \bibinfo {author} {\bibfnamefont {B.~P.}\ \bibnamefont
  {Flannery}},\ }\href {http://www.nr.com/} {\emph {\bibinfo {title} {Numerical
  Recipes 3rd Edition: The Art of Scientific Computing}}},\ \bibinfo {edition}
  {3rd}\ ed.\ (\bibinfo  {publisher} {Cambridge University Press},\ \bibinfo
  {address} {New York, NY, USA},\ \bibinfo {year} {2007})\BibitemShut {NoStop}%
\bibitem [{\citenamefont {White}\ and\ \citenamefont
  {Affleck}(2008)}]{white08}%
  \BibitemOpen
  \bibfield  {author} {\bibinfo {author} {\bibfnamefont {S.~R.}\ \bibnamefont
  {White}}\ and\ \bibinfo {author} {\bibfnamefont {I.}~\bibnamefont
  {Affleck}},\ }\href {\doibase 10.1103/physrevb.77.134437} {\bibfield
  {journal} {\bibinfo  {journal} {Phys. Rev. B}\ }\textbf {\bibinfo {volume}
  {77}},\ \bibinfo {pages} {134437} (\bibinfo {year} {2008})}\BibitemShut
  {NoStop}%
\bibitem [{\citenamefont {Barthel}\ \emph {et~al.}(2009)\citenamefont
  {Barthel}, \citenamefont {Schollw\"ock},\ and\ \citenamefont
  {White}}]{barthel09}%
  \BibitemOpen
  \bibfield  {author} {\bibinfo {author} {\bibfnamefont {T.}~\bibnamefont
  {Barthel}}, \bibinfo {author} {\bibfnamefont {U.}~\bibnamefont
  {Schollw\"ock}}, \ and\ \bibinfo {author} {\bibfnamefont {S.~R.}\
  \bibnamefont {White}},\ }\href {\doibase 10.1103/PhysRevB.79.245101}
  {\bibfield  {journal} {\bibinfo  {journal} {Phys. Rev. B}\ }\textbf {\bibinfo
  {volume} {79}},\ \bibinfo {pages} {245101} (\bibinfo {year}
  {2009})}\BibitemShut {NoStop}%
\bibitem [{\citenamefont {Anders}\ and\ \citenamefont
  {Schiller}(2005)}]{anders05}%
  \BibitemOpen
  \bibfield  {author} {\bibinfo {author} {\bibfnamefont {F.~B.}\ \bibnamefont
  {Anders}}\ and\ \bibinfo {author} {\bibfnamefont {A.}~\bibnamefont
  {Schiller}},\ }\href {\doibase 10.1103/physrevlett.95.196801} {\bibfield
  {journal} {\bibinfo  {journal} {Physical Review Letters}\ }\textbf {\bibinfo
  {volume} {95}},\ \bibinfo {pages} {196801} (\bibinfo {year}
  {2005})}\BibitemShut {NoStop}%
\bibitem [{\citenamefont {Nghiem}\ and\ \citenamefont
  {Costi}(2014)}]{nghiem14}%
  \BibitemOpen
  \bibfield  {author} {\bibinfo {author} {\bibfnamefont {H.~T.~M.}\
  \bibnamefont {Nghiem}}\ and\ \bibinfo {author} {\bibfnamefont {T.~A.}\
  \bibnamefont {Costi}},\ }\href {\doibase 10.1103/physrevb.90.035129}
  {\bibfield  {journal} {\bibinfo  {journal} {Phys. Rev. B}\ }\textbf {\bibinfo
  {volume} {90}},\ \bibinfo {pages} {035129} (\bibinfo {year}
  {2014})}\BibitemShut {NoStop}%
\bibitem [{\citenamefont {Wang}\ \emph {et~al.}(2014)\citenamefont {Wang},
  \citenamefont {Cohen},\ and\ \citenamefont {Xu}}]{wang14}%
  \BibitemOpen
  \bibfield  {author} {\bibinfo {author} {\bibfnamefont {P.}~\bibnamefont
  {Wang}}, \bibinfo {author} {\bibfnamefont {G.}~\bibnamefont {Cohen}}, \ and\
  \bibinfo {author} {\bibfnamefont {S.}~\bibnamefont {Xu}},\ }\href
  {http://arxiv.org/abs/1410.1480} {\bibfield  {journal} {\bibinfo  {journal}
  {preprint}\ } (\bibinfo {year} {2014})},\ \Eprint
  {http://arxiv.org/abs/1410.1480} {arXiv:1410.1480} \BibitemShut {NoStop}%
\end{thebibliography}
\end{document}